\documentclass[12pt]{article}
\usepackage{amsmath,amsthm,amsfonts,amssymb}
\usepackage{natbib}
\usepackage{fullpage} 

\expandafter\let\csname equation*\endcsname\relax
\expandafter\let\csname endequation*\endcsname\relax
\usepackage{psfrag}

\usepackage{graphicx,color}
\usepackage{epstopdf, epsfig}

\usepackage{enumerate}
\usepackage{empheq}

\def\eps{\varepsilon}
\def\d{{\mathrm d}}

\newcommand{\norm}[1]{\left\lVert#1\right\rVert}

\newcommand*{\grad}{\boldsymbol{\nabla}}
\renewcommand*{\u}{{\boldsymbol{u}}}
\renewcommand*{\v}{{\boldsymbol{v}}}
\newcommand*{\w}{{\boldsymbol{w}}}
\newcommand*{\x}{{\boldsymbol{x}}}
\newcommand*{\R}{{\boldsymbol{R}}}

\newcommand*{\F}{{\boldsymbol{F}}}
\renewcommand*{\a}{{\boldsymbol{a}}}
\renewcommand*{\k}{{\boldsymbol{k}}}
\newcommand*{\Eta}{{\boldsymbol{\eta}}}
\renewcommand*{\Xi}{{\boldsymbol{\xi}}}

\newcommand*{\Cdot}{\boldsymbol{\cdot}}

\newcommand*{\Bu}{\mathrm{Bu}}

\newcommand*{\ag}{{\operatorname{ag}}}

\newcommand*{\ini}{{\operatorname{in}}}
\newcommand*{\bal}{{\operatorname{bal}}}
\newcommand*{\hess}{\operatorname{Hess}}

\title{Comparison of variational balance models for the rotating shallow water equations}

\author{David G. Dritschel\thanks{School of Mathematics and Statistics, University of St Andrews, St Andrews KY16 9SS, UK, {\em{email}}: {david.dritschel@st-andrews.ac.uk}},
  Georg A. Gottwald\thanks{School of Mathematics and Statistics, University of Sydney, NSW 2006, Australia,  {\em{email}}:{georg.gottwald@sydney.edu.au}}
  and Marcel Oliver\thanks{School of Engineering and Science, Jacobs University, 28759 Bremen, Germany, {\em{email}}:{oliver@member.ams.org}}}

\begin{document}

\maketitle

\begin{abstract}
We present an extensive numerical comparison of a family of balance
models appropriate to the semi-geostrophic limit of the rotating
shallow water equations, and derived by variational asymptotics in
\citet{Oliver06} for small Rossby numbers ${\mathrm{Ro}}$. This family
of generalized large-scale semi-geostrophic (GLSG) models contains the
$L_1$-model introduced by \citet{Salmon83} as a special case. We use
these models to produce balanced initial states for the full shallow
water equations. We then numerically investigate how well these models
capture the dynamics of an initially balanced shallow water flow. It
is shown that, whereas the $L_1$-member of the GLSG family is able to
reproduce the balanced dynamics of the full shallow water equations on
time scales of ${\mathcal{O}}(1/{\mathrm{Ro}})$ very well, all other
members develop significant unphysical high wavenumber contributions
in the ageostrophic vorticity which spoil the dynamics.
\end{abstract}

{\bf{Keywords:}} rotating shallow water, semi-geostrophic equations, balance models

\section{Introduction}
\label{s.intro}

Atmospheric and oceanic large-scale flows are characterized by an
approximate balance between Coriolis forces, buoyancy and pressure
gradients.  This balance causes large-scale features such as the high
and low pressure fields which we experience as weather to vary only
slowly, and also implies that faster processes such as inertia-gravity
waves and acoustic waves are generally less important energetically.

Characterizing balance has been a longstanding problem in geophysical
fluid dynamics.  Four fundamental approaches are available.  First,
balance relations may be regarded as phase-space constraints in an
asymptotic expansion of the equations of motion, in a distinguished
limit of scaling parameters such as Rossby, Burger, and Froude number.
Second, similarly, asymptotic expansions may be performed on the
underlying Hamilton principle.  
Third, optimal balance strategies may be used to exploit the
adiabatic invariance of the slow or balanced manifold under
deformation \citep{ViudezDritschel04}.  Fourth, time-filters may
provide simple heuristics to distinguish balanced from imbalanced
motion.

While balance models are clearly not sufficient as a dynamical core
for contemporary weather or climate forecasting, there is continuing
necessity to characterize, diagnose, and enforce balance in the
context of such modeling.  Respecting balance has long been recognized
to be integral to the quality of weather forecasts. The first
numerical weather forecast (albeit performed with pen and paper) by
\citet{Richardson22} failed exactly because the initial fields used to
seed the forecast were imbalanced, containing an excessive amount of
small-scale high-frequency components, thereby spoiling the subsequent
forecast (see the wonderful historical account in \citet{Lynch}). In
modern weather forecasting, the initial state is estimated by
correcting the output from the forecast model, which may contain model
error as well as instabilities, using information from noisy
observations in a procedure called \emph{data assimilation}.  This
procedure, however, typically does not respect balance, with the
 consequence that it may produce highly imbalanced initial
states
\citep{BloomEtAl96,MitchellEtAl02,OurmieresEtAl06,Kepert09,GreybushEtAl11,Gottwald14}.


Within the vast literature on asymptotic derivations of balance
models, there are two main distinguished limits when the Rossby
number, the ratio of typical advective time scales to the time scale
of rotation, is small: (1) the quasi-geostrophic limit which assumes
that the Burger number (see Section~\ref{s.sw}) remains of order one
while variations in layer thickness are small, and (2) the
semi-geostrophic limit which assumes that the Burger number remains
small (comparable to the Rossby number) while variations in layer
thickness may be order one.  Quasi-geostrophy will not be considered
further in this paper.  The classic semi-geostrophic equations are
based on the geostrophic momentum approximation
\citep{eliassen1948quasi} and were rewritten by Hoskins and solved via
an ingenious change of coordinates
\citep{Hoskins75,cullen1984extended}.  They continue to attract
interest due to their connection to optimal transport theory and the
resulting possibility to make mathematical sense of generalized
frontal-type solutions \citep{benamou1998weak,cullen2008comparison}.
The geostrophic momentum approximation and Hoskins' transformation
inspired \cite{Salmon83,Salmon1985} to make corresponding
approximations directly to Hamilton's principle so as to preserve
geometrical structure and automatically preserve conservation laws.

In this paper, we perform a detailed numerical study of a particular
family of asymptotic balance models, based on the generalized
large-scale semi-geostrophic (GLSG) equations.  These equations
describe the motion of a rotating fluid in the limit of small Rossby
and small Froude number, and here for simplicity we consider a
single-layer shallow water flow only.  The GLSG equations go back to
an idea proposed by \cite{Salmon83,Salmon1985}.  He suggested imposing
a phase-space constraint directly in Hamilton's principle, that is, in
the \emph{variational} derivation of the model equations.
\cite{Oliver06} generalized this idea and derived a one-parameter
family of balance models, the GLSG family, that includes the two
models considered by Salmon as special cases.  Each member of this
family is characterized by a different choice of coordinates, and a
transformation from these new coordinates into physical coordinates is
given.  For exactly one of these models, namely Salmon's $L_1$-model,
this transformation is so close to the identity that physical
coordinates can be identified with model coordinates without changing
the asymptotic order of the model, as is implicitly done in Salmon's
work.

The GLSG equations can be formulated in terms of potential vorticity
advection and inversion, and can be shown to possess global smooth
solutions \citep{CalikEtAl13}. Moreover, this family of models is
distinct from other existing models in the semi-geostrophic limit such
as those derived in \cite{allen1996extended},
\cite{McintyreR2002HigherAA}, and the so-called $\delta$-$\gamma$
balance model hierarchy of \citet{MohebalhojehDritschel01}.  The
mathematical setting for this last group of models is less well
investigated and we shall not consider them further in this paper.

\cite{GottwaldO:2014:SlowDD} proved, in a structurally analogous
finite-dimensional context, that all models within the GLSG family
provide the same asymptotic order of accuracy.  In the
infinite-dimensional fluid dynamical model context, a corresponding
proof remains elusive.  
Moreover, it is already evident from an informal inspection of the resulting balance relations that the regularity provided by such relations for the constraint variables (or balanced variables) differs across the family of models.  
In fact, in
one of the cases considered by Salmon (the $L_1$-model which emerges
as the case $\lambda=\tfrac12$ in the notation introduced below), the
balance relation is an elliptic equation, in the other case
(corresponding to $\lambda=-\tfrac12$), the balance relation loses
ellipticity and the resulting model is ill-posed as an initial value
problem.  Moreover, \cite{Oliver06} showed that the balance relation
in a third special case (corresponding to $\lambda=0$) yields a
velocity field that is more regular by at least two derivatives than
any other member of the family.  It has thus been an obvious question
whether this apparent gain of regularity might be advantageous.

The main contribution of this paper is a careful comparative numerical
evaluation of the GLSG family of balance models.  Our main metric is a
comparison of the balance model dynamics to a consistently initialized
simulation of the full shallow water equations over a moderate
interval of time chosen such that the Eulerian fields experience an
order one relative change.  We examine in particular the scaling
behavior of the balance model error as the Rossby number goes to zero.
Our numerical and mathematical analyses underscore the importance in
understanding, and ensuring, the mathematical regularity of the
balance relation underpinning any balance model.  Crucially, our
results show that it is the regularity of the \emph{ageostrophic}
components of the flow that has the greatest impact on how well the
balance model is able to capture the dynamics of the full shallow
water model.  This singles out the $L_1$-model within the GLSG family
of balance models as the only model with sufficient regularity on the
ageostrophic vorticity to be viable in practice.

The paper is organized as follows.  Section~\ref{s.sw} presents the
shallow water equations and their semi-geostrophic
scaling. Section~\ref{s.oliver} briefly reviews the variational
asymptotics by \citet{Oliver06} and re-expresses the GSLG balance
relation in terms of ageostrophic variables.  Section~\ref{s.setup}
describes the setup of our numerical experiments, including the
details of the initialization procedure producing balanced initial
conditions for the rotating shallow water equations.
Section~\ref{s.numerics} present our results, showing that there
exists a distinguished balance model, namely Salmon's $L_1$-model,
which produces reliably balanced states which remain near balance over
times on which Eulerian fields evolve significantly.  We conclude with
a discussion and outlook in Section~\ref{s.summary}.


\section{The shallow water equations and the semi-geostrophic limit}
\label{s.sw}

The simplest model of rapidly rotating fluid flow in which the idea of
variational balance models can be tested is the rotating shallow
water model.  This describes the motion of a shallow layer of fluid
of mean height $H$, held down by gravity $g$ and rotating uniformly at
the rate $f/2$. The equations of motion (in the rotating frame of
reference) consist of the momentum equation and the continuity
equation, for the fluid velocity $\u=(u,v)$ and the height field $h$
(here, for convenience, scaled on $H$):
\begin{subequations}
  \label{e.shallow-dimensional}
  \begin{gather}
    \partial_t \u + \u \Cdot \grad \u 
      + f \u^\bot + c^2 \, \grad h = 0 \,,
    \label{e.shallow-dimensional.a} \\
    \partial_t h + \grad \Cdot (h\u) = 0 \,,
    \label{e.shallow-dimensional.b}
  \end{gather}
\end{subequations}
where $\u^\bot=(v,-u)$ and $c^2=gH$ is the squared short-scale gravity wave
speed.  For simplicity, we consider flow in the
doubly-periodic domain $\Omega=[0,2\pi]^2$. 

Notably, the above equations imply material conservation of
\emph{potential vorticity}
\begin{equation}
  q = \frac{f + \grad^\bot\Cdot\u}h ,
  \label{e.sg-sw}
\end{equation}
where $\grad^\bot \equiv (-\partial_y,\partial_x)$, that is
\begin{equation}
  \partial_t q + \u \Cdot \grad q = 0 \,.
  \label{e.PV_SW_dim}
\end{equation}
This is not an additional equation, but a consequence of combining
\eqref{e.shallow-dimensional.a} and \eqref{e.shallow-dimensional.b}.
Alternatively, conservation of potential vorticity can be derived as a
Noetherian conservation law from the particle-relabeling symmetry
\citep{Salmon}.

Under appropriate rescaling, 
the shallow water equations are characterized by several dimensionless
parameters. Taking $L$, $H$ and $U$ to be characteristic horizontal
length, height and velocity scales, respectively, two parameters
naturally emerge.  The first is the Rossby number
\begin{equation}
  \mathrm{Ro} = \frac{U}{fL}\, ,
\end{equation}
which measures the ratio of the relative vorticity $\zeta=\grad^\bot\Cdot\u$ of the fluid flow
to the planetary vorticity (or Coriolis frequency) $f$. In the
analysis below, we assume $\mathrm{Ro} \ll 1$. The second parameter is
the Froude number
\begin{equation}
  \mathrm{Fr} = \frac{U}{c} \,,
\end{equation}
which measures the flow speed relative to the characteristic gravity
wave speed $c=\sqrt{gH}$.  This too is considered small.  However, the
ratio of these small parameters determines the flow regime observed.
This is traditionally characterized by the Burger number
\begin{equation}
  \mathrm{Bu} = \frac{\mathrm{Ro}^2}{\mathrm{Fr}^2} = \frac{L_D^2}{L^2} \,,
  \label{e.Bu}
\end{equation}
where $L_D=c/f$, known as the Rossby radius of deformation,
signifies the length scale above which rotational effects dominate
over buoyancy effects.

Here we consider semi-geostrophic scaling, for which
$\Bu=\mathrm{Ro}$, in contrast to the more extensively studied
quasi-geostrophic scaling, for which $\Bu={\mathcal{O}}(1)$ and height
perturbations are of $\mathcal{O}(\mathrm{Ro})$ to maintain
geostrophic balance at leading order.  Notably, in semi-geostrophic
scaling, (rescaled) height variations may be ${\mathcal{O}}(1)$.

The nondimensional equations are found by scaling $x$ and $y$ by $L$,
$\u$ by $U$, $h$ by a mean height $H$, and $t$ by $L/U$.  Defining
$\eps \equiv \mathrm{Ro}\ll 1$ as our small parameter and setting
$\Bu=\eps$, the equations become
\begin{subequations}
  \label{e.shallow}
  \begin{gather}
    \eps \, (\partial_t \u + \u \Cdot \grad \u) 
      + \u^\bot + \grad h = 0 \,,
    \label{e.shallow.a} \\
    \partial_t h + \grad \Cdot (h\u) = 0 \,.
    \label{e.shallow.b}
  \end{gather}
\end{subequations}
The non-dimensional potential vorticity is
\begin{equation}
  q = \frac{1 + \eps \, \grad^\bot\Cdot\u}h \,.
  \label{e.sg-sw-dim}
\end{equation}

In the derivation of the GLSG balance models below, we will use the
above non-dimensional form of the equations.  
Note, due to the rescaling adopted, the mean height $\bar h\equiv 1$.


\section{A family of balance models}
\label{s.oliver}

In this section, we give a brief review of the family of first-order
generalized Lagrangian semi-geostrophic (GLSG) models which were derived
in \cite{Oliver06}. These models are asymptotic models for small
Rossby number under semi-geostrophic scaling. Rather than performing
asymptotics directly on the equations of motion \eqref{e.shallow},
\citet{Oliver06} followed the strategy of \citet{Salmon83,Salmon} and
performed the asymptotics within the variational principle, thereby
guaranteeing the conservation of the geometric Hamiltonian structure
of the original shallow water equations.

\subsection{A variational principle for shallow water}

It is well known that the shallow water equations arise as the Euler--Lagrange equations from a variational principle; see, for example, \cite{Salmon83,Salmon}.  In our opinion, it is most clearly presented using the following notation.  We consistently write $\x$ to denote an Eulerian position and $\a$ to denote a Lagrangian label of a fluid parcel.  The \emph{flow map} $\Eta$ maps labels to Eulerian positions such that the parcel initially at location $\a$ is at location $\x = \Eta(\a,t)$ at time $t$.  We write $\u = \u(\x,t)$ to denote the (Eulerian) velocity of the fluid at location $\x$ and time $t$.  It equals the (Lagrangian) velocity of the parcel passing through $\x$ at time $t$, so that $\partial_t \Eta (\a,t) = \u(\Eta(\a,t),t)$, which we shall abbreviate
\begin{equation}
  \label{e.u-flow}
  \dot \Eta = \u \circ \Eta \,,
\end{equation} 
the symbol ``$\circ$'' denoting composition of maps with respect to
the spatial variables.  Liouville's theorem states that the
continuity equation \eqref{e.shallow.b} is equivalent to
\begin{equation}
  \label{e.h}
  h \circ \Eta = \frac{h^\ini}{\det \grad\Eta} \,,
\end{equation}
where $h^\ini = h^\ini(\a)$ is the initial height field.  To simplify
the derivation of the equations of motion, we suppose for a moment
that $h^\ini=1$.  This can be done without loss of generality because
the equations of motion do not depend on the choice of the initial
height field.  With this convention, the layer depth is the Jacobian
of the transformation from Eulerian to Lagrangian coordinates.

We can now introduce the Lagrangian
\begin{align}
  L & = \int
        \bigl[
          (\R + \tfrac12 \, \varepsilon \, \u) \circ \Eta
          \Cdot \dot \Eta - \frac{1}{2} \, h \circ \Eta
        \bigr] \, \d\a \notag \\
    & = \int h \,
        \bigl[
          \R \Cdot \u
          + \tfrac12 \, \varepsilon \, \lvert \u \rvert^2
          - \frac{1}{2} \, h
        \bigr] \, \d\x \,,
        \label{e.LagrSHW}
\end{align}
where $\R$ denotes the vector potential corresponding to the Coriolis
parameter, such that $\grad^\bot \Cdot \R = f \equiv 1$.  It is not
hard to show that the shallow water equations \eqref{e.shallow} arise
as the stationary points of the \emph{action}
\begin{equation}
  S = \int_{t_1}^{t_2} L[\u,h] \, \d t 
\end{equation}
with respect to variations of the flow map $\Eta$.  Since $h$ and $\u$ are linked to $\Eta$ by relations \eqref{e.u-flow} and \eqref{e.h} above, variations in $\Eta$ induce variations in $h$ and $\u$ of a specific form.  This is most easily expressed by noting that a variation of the flow map $\delta \Eta$ can be thought of as induced by an Eulerian vector field $\w=\w(\x)$ via
\begin{equation}
  \label{e.w-flow}
  \delta \Eta = \w \circ \Eta \,,
\end{equation}
a direct analog to relation \eqref{e.u-flow}.  Just the same, there holds a Liouville theorem with respect to a parametrization of the variation, which implies the ``continuity equation''
\begin{equation}
  \label{e.delta-h}
  \delta h + \grad \Cdot (\w h) = 0 \,.
\end{equation}
Finally, cross-differentiation of \eqref{e.u-flow} and
\eqref{e.w-flow} yields the so-called \emph{Lin constraint}
\citep{Bretherton70}:
\begin{equation}
  \label{e.delta-u}
  \delta \u = \dot \w + \grad \w \, \u - \grad \u \, \w \,.
\end{equation}
The remainder of the derivation proceeds by direct computation and
shall be omitted.  We remark, however, that the argument requires that
the Coriolis parameter $f$ can be written as the curl of a vector
potential.  On the plane, this is easy to achieve.  However, on the
torus, $f$ has a vector potential if and only if it has zero mean,
thereby excluding the case of a constant Coriolis parameter considered
here.  However, a careful inspection of the problem shows that if we
proceed as if the vector potential $\R$ existed, we would obtain
equations of motion which are Hamiltonian in the expected sense, but
strictly speaking do not arise as the Euler--Lagrange equations of a
variational principle.  A detailed discussion of this issue is given
in \cite{OliverVasylkevych11}.  We shall henceforth ignore this
subtlety as it is not pertinent to the main point of this paper.

\subsection{Variational asymptotics}

The key idea introduced in \cite{Oliver06} is to introduce a new
coordinate system which is related to the original coordinate system
by an ${\mathcal{O}}(\eps)$ perturbation of the identity in such a way that the
first-order transformed Lagrangian becomes degenerate.  As a result,
truncation of the Lagrangian to first order leads to Euler--Lagrange
equations which live on a reduced phase space.

To systematically develop this idea, it is convenient to view the
transformation itself as a flow parametrized by $\eps$.  Concretely,
we shall endow quantities in the original (physical) coordinate system
with a subscript $\eps$, while quantities without subscript shall be
viewed as posed in a new, slightly distorted coordinate system.  (This
choice makes the transformation from new to old coordinates explicit,
while the transformation from old to new coordinates is implicit.)  
We
view the transformation as being generated by a vector field $\v_\eps$
via
\begin{equation}
  \label{e.v-eps}
  \Eta_\varepsilon' = \v_\varepsilon \circ \Eta_\varepsilon \,,
\end{equation}
where the prime denotes a derivative with respect to $\varepsilon$ and
$\Eta_0 \equiv \Eta$.  Once more, we have a continuity equation which
now reads
\begin{equation}
  h_\eps' + \grad \Cdot (\v_\eps h_\eps) = 0\; ,
  \label{e.hp}
\end{equation}
and an analog of the Lin constraint \eqref{e.delta-u},
\begin{equation}
  \u_\eps' = \dot \v_\eps + \grad \v_\eps \, \u_\eps
             - \grad \u_\eps \, \v_\eps \,.
\label{e.up}
\end{equation}
In the following, we shall denote the formal Taylor coefficients of
$\u_\eps$ with respect to an expansion in $\eps$ by $\u$, $\u'$, etc.,
with analogous notation for all other quantities.

In summary, altogether we are considering a three parameter family of
flow maps, the parameters being physical time $t$, asymptotic
parameter $\eps$, and an implicit parameter in the definition of the
variational derivative.  Structurally, these parameters play entirely
symmetric roles, the difference lies in their physical interpretation.
Each Lagrangian-parameter derivative has an associated Eulerian vector
field: $\u_\eps$ for the time derivative, $\v_\eps$ for the
$\eps$-derivative, and $\w_\eps$ for the variational derivative.  We
can interpret $\v_\eps$ as the velocity of deformation of the
coordinate system in ``artificial time'' $\eps$, and $\w_\eps$ as the
Eulerian version of the virtual displacement in classical mechanics
(e.g. \citealt{Goldstein}).  The definition of $h_\eps$ as the inverse
Jacobian of the map $\Eta_\eps$ implies a continuity equation in each
of these parameters, stated in \eqref{e.shallow.b}, \eqref{e.hp}, and
\eqref{e.delta-h}, respectively.  Mixed derivatives satisfy
generalized Lin constraints such as \eqref{e.delta-u} and
\eqref{e.up}.

We now proceed to expand the Lagrangian \eqref{e.LagrSHW} in powers of
$\eps$:
\begin{equation}
  L_\varepsilon 
  = \int \bigl[
            \R \circ \Eta \Cdot \dot \Eta
            - \frac{1}{2} \, h \circ \Eta
          \bigr] \, \d\a
    + \eps \, \int
          \bigl[
            \v^\bot \Cdot \u
            + \tfrac12 \, \lvert \u \rvert^2
            + \frac12 \, h \, \grad \Cdot \v
          \bigr] \circ \Eta \, \d\a 
    + {\mathcal{O}}(\eps^2) \,.      
  \label{e.l-expand}
\end{equation}
Details of this calculation can be found in Appendix~B of
\cite{Oliver06}.  The transformation vector field $\v$ at
${\mathcal{O}}(\eps)$ may be chosen arbitrarily.  Clearly, any choice
of the form
\begin{equation}
  \v = \tfrac12 \, \u^\bot + \F(h) 
\end{equation}
renders the first-order Lagrangian $L_1$ affine (i.e., it is linear in
the velocity and thus degenerate).  The dimensionally consistent
choice
\begin{equation}
  \v = \tfrac12 \, \u^\bot + \lambda \, \grad h 
  \label{e.v}
\end{equation}
leads to a particular one-parameter family of balance models --- when
the system is in geostrophic balance, the second term is a scalar
multiple of the first.  In this setting, the choice $\lambda=\tfrac12$
emerges as a special case: at leading order, both terms cancel so
that, formally, $\v= \mathcal O(\eps)$. 

Inserting the choice \eqref{e.v} back into \eqref{e.l-expand} and
dropping terms of order ${\mathcal{O}}(\eps^2)$, we obtain
\begin{equation}  
  \label{e.l-bal}
  L_\bal = \int \bigl[
                  \R + 
                  \eps \, (\lambda + \tfrac12) \, \grad^\bot h 
                \bigr] \circ \Eta \Cdot \dot \Eta \, \d\a
         - \int h \, 
                \bigl[
                  \tfrac12 \, h + 
                  \eps \, \lambda \, \lvert \grad h \rvert^2
                \bigr] \, \d\x \,,
\end{equation}
where, for convenience, we have written the part which is linear in
$\u$ as an integral over labels and the part which only depends on $h$
as an integral over Eulerian positions.

For the convenience of the reader, the explicit variational calculus
of $L_\bal$ is presented in Appendix~\ref{a.oliver}. The stationary
points of the action necessitate the Euler--Lagrange equation
\begin{equation}
  \bigl[
    1 - \eps \, (\lambda + \tfrac12) \,
        (h \, \Delta + 2 \, \grad h \Cdot \grad)
  \bigr] \u
  = \grad^\bot
  \bigl[
    h - \eps  \, \lambda \,
    (2 \, h \, \Delta h + \lvert \grad h \rvert^2)
  \bigr] \,,
  \label{e.elliptic}
\end{equation}
where $\Delta$ denotes Laplace's operator.  For a given nonnegative
height field $h$, this equation is a non-constant coefficient elliptic
equation for $\u$ when $\lambda > -\frac12$. This constitutes the
family of balance relations, parametrized by the free parameter
$\lambda$.  

The system of equations for the balance model can be closed via the
continuity equation
\begin{equation}
  \partial_t h + \grad \Cdot (h \u) = 0 \,.
  \label{e.cont}
\end{equation}
By construction, the balance model has a conserved energy,
\begin{equation}
  H_\bal 
  = \frac12 \int h^2 \, \d \x
    + \eps \, \lambda \int h \, \lvert \grad h \rvert^2 \, \d \x \,,
\end{equation}
and a materially conserved \emph{potential vorticity}
\begin{equation}
  q = \frac{1 + \eps \, (\lambda + \frac12) \, \Delta h}h \,.
  \label{e.sg-pv}
\end{equation}
Thus, we can choose either \eqref{e.cont} or the potential vorticity
conservation law
\begin{equation}
  \partial_t q + \u \Cdot \grad q = 0 
  \label{e.PV}
\end{equation}
to evolve the balance relation \eqref{e.elliptic} in time.  This
equivalence can be checked by brute-force computation, or by noting
that potential vorticity advection is the natural conservation law
associated with the particle relabeling symmetry in the variational
derivation of the balanced models \citep{Oliver06}.  If we opt for $q$
as the fundamental prognostic variable, the height field $h$ can be
recovered by inversion of \eqref{e.sg-pv} via
\begin{align}
  \bigl( q - \eps \, (\lambda + \tfrac{1}{2}) \Delta \bigr) h 
  = 1 \,,
  \label{e.sg-pv-inv}
\end{align}
after which $\u$ is computed from $h$ via \eqref{e.elliptic}.  Thus,
\eqref{e.PV}, \eqref{e.sg-pv-inv}, and \eqref{e.elliptic} form a
closed system for the balanced dynamics.  This formulation is used
numerically and also underlies the proof of global well-posedness
\citep{CalikEtAl13} and of global existence of weak solutions
\citep{CalikOliver13} for the family of balance models.

Note that, to leading order, the motion induced by a velocity field
computed from \eqref{e.elliptic} is geostrophic with an
${\mathcal{O}}(1)$ velocity.  Thus, fluid parcels travel a unit
distance over times of ${\mathcal{O}}(1)$.  The rate of change of the
Eulerian fields, on the other hand, is determined by the magnitude of
the ageostrophic velocity which is ${\mathcal{O}}(\eps)$, independent
of $\lambda$.  (An explicit formal calculation can be found in
Appendix~\ref{a.time-scale}.)  Thus, to test the prognostic skill of
the balance model, we need to simulate on time scales of order
${\mathcal{O}}(\eps^{-1})$.

\subsection{Balance relation in $\delta$-$\gamma$ variables}
\label{ageosection}

For the understanding of the behavior of the balance relation for
different values of the free parameter $\lambda$, it is crucial to look
at the effect of the balance relation
on the \emph{ageostrophic velocity} in balance model
coordinates,
\begin{equation}
  \u^\ag = \u - \grad^\bot h \,.
\end{equation}
We choose to re-express the ageostrophic motion in terms of the
balance model \emph{divergence} $\delta = \grad \Cdot \u$ and
\emph{ageostrophic vorticity}
$\gamma = \grad^\bot \Cdot \u^\ag = \grad^\bot \Cdot \u - \Delta h$.
Strictly speaking, the ageostrophic vorticity is better described as
the \emph{acceleration divergence} since, for the full shallow water
equations, $\gamma = \grad \Cdot(\partial_t \u + \u \Cdot \grad \u)$
via the shallow water momentum equation and this characterization
remains appropriate in spherical geometry \citep{SmithDritschel:2006}.
Nevertheless, in the
following we shall refer to $\gamma$ as the ageostrophic vorticity.

We now rewrite the balance relation in terms of $\delta$ and $\gamma$
by taking the divergence and curl of \eqref{e.elliptic}, obtaining
\begin{subequations}
  \label{e.ageo-balance}
\begin{align}
  \bigl[
    1 - \eps \, (\lambda + \tfrac12) \,
    (h \, \Delta + 2 \, \grad h \Cdot \grad)
  \bigr] \delta 
  = \eps \, (\lambda + \tfrac12) \, 
    (\grad h \Cdot \Delta \u 
    + 2 \, \grad \grad h : \grad \u) 
  \label{e.delta_eps} 
\end{align}
and
\begin{multline}
    \bigl[
    1 - \eps \, (\lambda + \tfrac12) \,
        (h \, \Delta + 2 \, \grad h \Cdot \grad)
  \bigr] \gamma
  = \eps \, (\lambda + \tfrac12) \, 
      (\grad^\bot h \Cdot \Delta \u 
       + 2 \, \grad \grad^\bot h : \grad \u) \\
    + \eps \, (\tfrac12 - \lambda) \, h \, \Delta^2 h
    + \eps \, (1-4 \lambda) \, \grad h \Cdot \grad \Delta h
    - 2 \eps \lambda \, ((\Delta h)^2 + \lvert \grad \grad h \rvert^2) \,,  
     \label{e.ageovort_eps} 
\end{multline}
\end{subequations}
where $A:B$ denotes the matrix inner product
$A:B = \sum_{i,j}a_{ij} \, b_{ij}$.  To eliminate all references to
$\u$ on the right hand sides, we decompose $\u$ into its rotational
and divergent components by writing
\begin{equation}
  \u = \grad^\bot \psi + \grad \phi
\end{equation}
so that
\begin{equation}
  \psi = \Delta^{-1} \gamma + h
  \qquad \text{and} \qquad
  \phi = \Delta^{-1} \delta \,.
\end{equation}
Then
\begin{equation}
  \grad^\bot h \Cdot \Delta \u 
  = \grad h \Cdot \Delta \grad h + \grad h \Cdot \grad \gamma
    + \grad^\bot h \Cdot \grad \delta
  \label{e.subs1}
\end{equation}
and
\begin{equation}
  \grad \grad^\bot h : \grad \u
  = \lvert \grad \grad h \rvert^2 
    + \grad \grad h : \grad \grad \Delta^{-1} \gamma
    + \grad \grad^\bot h : \grad \grad \Delta^{-1} \delta \,,
  \label{e.subs2}
\end{equation}
with similar relations for the terms on the right hand side of
\eqref{e.delta_eps}.  Inserting these expressions back into
\eqref{e.ageo-balance} and rearranging terms, we obtain
\begin{subequations}

\begin{align}
    \bigl(1 & - \eps \, (\lambda + \tfrac12) \,
    h \, \Delta \bigr) \delta  
  = \eps \, (\lambda + \tfrac12) \, \grad h \Cdot \grad^\bot \Delta h
    \notag \\
  & \quad + \eps \, (\lambda + \tfrac12) \, 
    \bigl(
      \grad h \Cdot \grad^\bot \gamma 
      + 2 \, \grad \grad h : \grad \grad^\bot \Delta^{-1} \gamma
      + 3 \, \grad h \Cdot \grad \delta
      + 2 \, \grad \grad h : \grad \grad \Delta^{-1} \delta 
    \bigr) 
  \label{e.delta_eps2} 
\end{align}
and
\begin{align}
  \bigl(1 & - \eps \, (\lambda + \tfrac12) \,
        h \, \Delta \bigr) \gamma
        = - 2 \eps \, \det \hess h 
        \notag \\
  & \quad + \eps \, (\tfrac12 + \lambda) \, 
      \bigl(
        3 \, \grad h \Cdot \grad \gamma
        + 2 \, \grad \grad h : \grad \grad \Delta^{-1} \gamma
        + \grad^\bot h \Cdot \grad \delta
        + 2 \, \grad \grad^\bot h : \grad \grad \Delta^{-1} \delta
      \bigr) 
        \notag \\
  & \quad + \eps \, (\tfrac12 - \lambda) \, 
      \bigl(
        h \, \Delta^2 h + 3 \, \grad h \Cdot \grad \Delta h
        + 2 \, (\Delta h)^2
      \bigr) \,.
  \label{e.ageovort_eps2} 
\end{align}
\end{subequations}
The operator on the left-hand sides is elliptic for $\lambda>-\tfrac12$. 
The terms in each of the second lines are linear in $\delta$ or
$\gamma$, hence are formally of ${\mathcal{O}}(\eps^2)$.  Thus, at
least when $\lambda=\tfrac12$, the dominant contribution comes from
the first term on each right hand side.

However, when $\lambda\neq \tfrac12$, the right hand side of balance
relation \eqref{e.ageovort_eps2} features additional terms involving
third and fourth-order derivatives of $h$, alongside second-order
derivatives.  Thus, the regularity of the ageostrophic vorticity is
severely reduced unless $\lambda=\tfrac12$.  Our numerical results
show that this loss of regularity, which affects the balance relation
for $\gamma$ only, has significant detrimental effects on the
prognostic skill of the balance model, as discussed in detail in
Sections~\ref{s.lambdachoice}--\ref{s.powerspec} below.

Our numerical results further show that the dominant right hand term
in the balance relation for $\delta$ \eqref{e.delta_eps2}, namely
$\grad h \Cdot \grad^\bot \Delta h$ shown as the blue curve in the
bottom row of Figure~\ref{f.fig9}, appears more regular than the the
corresponding term in the balance relation for $\gamma$
\eqref{e.ageovort_eps2}, namely $\grad h \Cdot \grad \Delta h$ shown
as the magenta curve on the top row of Figure~\ref{f.fig9}.  The cause
of the apparent cancellations in the former term is currently not
understood.

\subsection{Transformation to shallow water coordinates}
\label{s.trafo}

Since the balance model dynamics, for each of the models introduced
above, is posed in a coordinate frame different from the physical
frame of the full shallow water dynamics, we need to apply a
coordinate transformation for consistent initialization and
diagnostics.  The transformation between the two is explicit going
from model coordinates to physical coordinates.  Its inverse is
defined implicitly and will generally involve an infinite series in
$\eps$.  Therefore, except for the case of the $L_1$ model, it is not
possible to write out the balance model in physical coordinates.

For consistent initialization and diagnostics of our numerical tests,
we need to write out the change of coordinates explicitly.  As we are
only carrying terms to ${\mathcal{O}}(\eps)$, we have
\begin{align}
  h_\eps = h + \eps \, h^\prime 
  \qquad \text{and} \qquad
  \u_\eps = \u + \eps \u^\prime
  \label{e.trafo}
\end{align}
with
\begin{subequations}
\begin{gather}
  h' = - \grad \Cdot (\v h) \,, \\
  \u^\prime = \dot \v + \grad \v \, \u - \grad \u \, \v \,,
\end{gather}
\end{subequations}
where
\begin{align}             
  \v = \tfrac12 \, \u^\bot + \lambda \, \grad h \,,
\end{align}
and where $\u$ and $h$ satisfy the balance relation \eqref{e.elliptic}.  We
then compute the shallow water potential vorticity, divergence, and
ageostrophic vorticity via
\begin{gather}
  q_\eps = (1 + \grad^\bot \cdot \u_\eps)/h_\eps \,, \qquad
  \delta_\eps = \grad \Cdot \u_\eps \,, 
  \qquad \text{and} \qquad
  \gamma_\eps = \grad^\bot \Cdot \u_\eps - \Delta h_\eps \,.
  \label{e.transformed-fields}
\end{gather}
When presenting our results, we will use the suggestive notation
$\operatorname{T}[q]$, $\operatorname{T}[\delta]$, and
$\operatorname{T}[\gamma]$ for the fields obtained via transformation
from the balance model quantities, with the understanding that this
notation does not imply any strict functional dependence --- all of
these transformed quantities are functionally dependent only on the
balance model potential vorticity $q$.

The transformation from physical coordinates to balance model
coordinates cannot be written down explicitly.  However, it is
possible to numerically invert the transformation for moderate values
of the characteristic parameters using an iterative scheme sketched in
Appendix~\ref{a.dynbal}.

We finally remark that the transformation involves taking time
derivatives of $\u$ and $h$.  Formally, these terms are
${\mathcal{O}}(\eps)$ as verified in Appendix~\ref{a.time-scale}.
Moreover, when $\lambda=\tfrac12$, then $\v$ itself is
${\mathcal{O}}(\eps)$ and the transformation remains
${\mathcal{O}}(\eps^2)$ --- and thus coincides with the identity up to
the formal order of validity of the balance model. Practically, this
means that the transformation can be omitted when $\lambda=\tfrac12$,
i.e.\ the fields of the balanced equations and of the full shallow
water equations can be directly compared without affecting the formal
order of accuracy.  We have numerically verified that the effect of
the transformation is indeed negligible for this particular case; our
detailed results, however, are computed with the transformation
applied for all values of $\lambda$.

We stress that the GLSG balance models consist of \emph{both} the
prognostic equation \eqref{e.elliptic} with assocated potential
vorticity inversion \eqref{e.PV} and \eqref{e.sg-pv-inv}, \emph{and}
the near-identity transformation relating the balance model solution
to the corresponding quantities in a physical coordinate frame.  When
transformed back to physical coordinates, all the models considered
here have the same $\mathcal{O}(\eps)$ order of accuracy at least
formally, the only difference being that the transformation is
necessary to maintain order when $\lambda \neq \tfrac12$.  In finite
dimensions, this statement is rigorous \citep{GottwaldO:2014:SlowDD}.
In the present setting, loss of accuracy can only be due to analytical
issues in infinite dimensions, not due to an inconsistent handling of
terms in the formal expansion.  We note, in particular, that the
transformed balance model potential vorticity given by
\eqref{e.transformed-fields} coincides with the shallow water
potential vorticity \eqref{e.sg-sw-dim} up to terms of
$\mathcal{O}(\eps^2)$ for all values of $\lambda$.  Similarly, the
balanced model Hamiltonian, transformed back to physical coordinates,
coincides with the shallow water Hamiltonian up to terms of
$\mathcal{O}(\eps^2)$ for all values of $\lambda$.

\section{Experimental setup}
\label{s.setup}

\subsection{Benchmarking scheme}

We now describe the details of our benchmarking procedure to determine
how well the different GLSG balance models are able to describe nearly
balanced shallow water dynamics.  For a fixed value of the parameter
$\lambda$, we go through the following steps:
\begin{enumerate}[{Step}~1:\;\;\;]
\item At time $t=0$, specify the initial balance model height field
$h^\ini$.
\item On the balance model side, compute the initial balance model
potential vorticity $q^\ini$ using \eqref{e.sg-pv}.
\item Compute the initial shallow water
$q_\eps^\ini = \operatorname{T}[q^\ini]$,
$\delta_\eps^\ini = \operatorname{T}[\delta^\ini]$, and
$\gamma_\eps^\ini = \operatorname{T}[\gamma^\ini]$ via the equations
detailed in Section~\ref{s.trafo}.
\item Evolve the balance model potential vorticity $q$ to some final
state $q(\x,t)$ at time $t$ using \eqref{e.PV}.  The balance model height
field $h$ and velocity field $\u$ are kinematically slaved to $q$ via
\eqref{e.sg-pv-inv} and \eqref{e.elliptic}, respectively, and computed
as part of the forward evolution.
\item Evolve the shallow water fields $q_\eps$, $\delta_\eps$, and
$\gamma_\eps$ to the same final time $t=1/\eps$.
\item Transform the balance model state, at any chosen time, to shallow water
coordinates as detailed in Section~\ref{s.trafo}; 
compare $q_\eps$ with $\operatorname{T}[q]$, $\delta_\eps$ with
$\operatorname{T}[\delta]$, and $\gamma_\eps$ with
$\operatorname{T}[\gamma]$.
\end{enumerate}
It is also possible to initialize on the shallow water side, i.e.\
given only the initial  distribution of shallow water potential
vorticity $q_\eps^\ini$, see Appendix~\ref{a.dynbal}.  This is more
demanding computationally, but does not affect any of our conclusions.
Such an initialization may be useful for quantifying the amount of
imbalance (or departure from balance) occurring over the course of a
shallow water simulation.  This however is not the aim of the present
study; instead we seek to determine how well balance models can
predict a shallow water flow evolution.

\subsection{The shallow water equations in $q$-$\delta$-$\gamma$
coordinates}

The shallow water model requires additional care since in this case
there are three prognostic variables, not one as in the balance model.
In the shallow water model we employ potential vorticity $q_\eps$,
divergence $\delta_\eps$ and ageostrophic vorticity $\gamma_\eps$
rather than more traditional choices like $h_\eps$, $u_\eps$, and
$v_\eps$, or like $h_\eps$, $\zeta_\eps$, and $\delta_\eps$.  Previous
work has shown that $q_\eps$, $\delta_\eps$, and $\gamma_\eps$ offer
significant advantages over traditional variable choices
\citep{MohebalhojehDritschel01,MohebalhojehDritschel04}. In
particular, they offer significantly greater accuracy in the
representation of \emph{both} the balanced and imbalanced parts of the
flow.  Moreover, $q_\eps$, $\delta_\eps$, and $\gamma_\eps$ lead to
linear elliptic problems to determine $h_\eps$ and $\u_\eps$,
advantageous for both numerical robustness and efficiency.

Ignoring hyperviscosity, the prognostic equations for $\delta_\eps$ and
$\gamma_\eps$ (in dimensional terms) are
\begin{subequations}
\begin{gather}
  \partial_t\delta_\eps = \gamma_\eps+2 \, J(u_\eps,v_\eps)
  -\grad\Cdot(\u_\eps\delta_\eps)\, , 
   \label{e.deltaeps}
  \\
  \partial_t\gamma_\eps = -f^2 \, \delta_\eps
  +c^2 \, \Delta\grad\Cdot(\u_\eps h_\eps)
  -f \, \grad\Cdot(\u_\eps\zeta_\eps)\, ,
  \label{e.gammaeps}
\end{gather}
\end{subequations}
where
$J(a,b) = \partial_x a \, \partial_y b-\partial_y a \, \partial_x b$,
and $\zeta_\eps = h_\eps q_\eps - f$ is the relative vorticity.  

The fields $\delta_\eps$, $\gamma_\eps$, and $q_\eps$ determine the
velocity field $\u_\eps$ only up to a spatially independent mean flow
$\bar{\u}_\eps(t)$.  In general, this flow is non-zero, though
typically of very small amplitude (we have checked that in the
numerical simulations presented below the mean flow has an amplitude
of about $10^{-4}\, \lvert \u_\eps \rvert_{\rm{max}}$).  It is taken
into account not only for completeness but to ensure an accurate
assessment of the differences between the shallow water and the
transformed balance flow solutions.  To write out the evolution
equation for $\bar{\u}_\eps(t)$, we take the average of
\eqref{e.shallow-dimensional.a}:
\begin{equation}
  \partial_t \bar \u_\eps
  = - \overline{(f + \zeta_\eps) \, \u_\eps^\bot}
  = - \overline{h_\eps q_\eps \u_\eps^\bot} \,.
\end{equation}
These additional two ordinary differential equations complete the set
of prognostic equations of the $q$-$\delta$-$\gamma$ formulation of
the shallow water equations.  The initial mean flow is determined as
the spatial average of the initial velocity field which is available
via the transformation from balance model coordinates.

From $\delta_\eps$, $\gamma_\eps$, and $q_\eps$, the fields $h_\eps$
and $\hat \u_\eps$, the mean-free component of $\u_\eps$, are
recovered by linear inversion.  First, the definition
$\gamma_\eps = f \, \zeta_\eps-c^2 \, \Delta h_\eps$, the definition
of $\zeta_\eps$, and the normalization of the mean height
$\bar h_\eps \equiv 1$ (see Section~\ref{s.sw}) lead to
\begin{equation}
  c^2 \, \Delta \hat h_\eps - f \, q_\eps \hat h_\eps 
  = - \gamma_\eps - f^2 + f \, q_\eps \,,
  \label{e.heps_ell}
\end{equation}
a linear elliptic equation for $\hat h_\eps$, the mean-free component
of $h_\eps$.
Then, once $h_\eps = \hat h_\eps + 1$ is determined, $\hat \u_\eps$ is
simply found using the Helmholtz decomposition
$\hat \u_\eps=\grad^\perp\psi_\eps+\grad\phi_\eps$.  This results in
the Poisson equations $\Delta\psi_\eps=\zeta_\eps$ and
$\Delta\phi_\eps=\delta_\eps$, both of which are directly solved in
spectral space.  The velocity $\hat \u_\eps$ is then found by
differentiation of $\psi_\eps$ and $\phi_\eps$, and $\u_\eps = \hat
\u_\eps + \bar \u_\eps$.

\subsection{Implementation}

The numerical models developed for the shallow water and balance
equations, including all initialization and diagnostic procedures,
make use of the standard pseudo-spectral method in doubly-periodic
geometry.  In this method, nonlinear products are carried out in
physical space (on a regular grid), while all linear operations such
as differentiation and inversion are carried out in spectral space.
Fast Fourier transforms are used to go from one representation to the
other.

To minimize aliasing errors, prior to carrying out nonlinear products,
fields are spectrally truncated using a circular filter of radius
(wavenumber magnitude) $k=n_{\rm g}/3$ where $n_{\rm g}$ is
the grid resolution in both $x$ and $y$ (here the domain is square
with side length $2\pi$ without loss of generality).  Note, the
maximum wavenumber is $k_{\rm{max}}=n_{\rm g}/2$.  We use
$n_{\rm g}=256$ throughout but have verified that
$n_{\rm g}=512$ does not change the results significantly in a
sample of cases.  While proper de-aliasing would remove more modes,
the circular filter adopted better preserves isotropy and has been
found to be sufficient to avoid noticeable aliasing errors.

The flow evolution models employ a standard fourth-order Runge--Kutta
time stepping method, with an adaptive time step.  The time step
$\Delta t$ is required to be simultaneously smaller than
$\Delta t_{\rm{gw}}$, $\Delta t_{\rm{cfl}}$, and
$\Delta t_{\rm{\zeta}}$; here $\Delta t_{\rm{gw}}=\Delta x/c$
is the gravity-wave resolving time step, while
$\Delta t_{\rm{cfl}}=0.7 \, \Delta x/|\u|_{\rm{max}}$ is the
CFL time step (with CFL parameter $0.7$), and
$\Delta t_{\rm{\zeta}}=\pi/(10 \, |\zeta|_{\rm{max}})$, where
$\zeta=\grad^\bot\Cdot\u$ is the relative vorticity.  In practice,
$\Delta t_{\rm{gw}}$ is always the smallest, so the time step is
fixed.

The flow evolution models also employ weak hyperviscosity, of the form
$\nu_{\rm{hyp}} \, \Delta^3 a$, for all evolved fields $a$.
Spectrally, this corresponds to subtracting
$r \, (k/k_{\rm{max}})^6 \, a_\k$ from the right-hand-side of the
evolution equation for the Fourier coefficients $a_\k$ of each field
$a$.  In the numerical implementation, this term is incorporated in
the time-stepping method exactly through an integrating factor.  The
damping rate $r$ on the highest wavenumber is chosen as
$r=10 \, \eps^2f$, after careful experimentation.  In practice, over
the moderate integration times carried out, the effect of
hyperviscosity is negligible.

To numerically determine the height field $\hat h_\eps$, we employ the
elliptic diagnostic equation \eqref{e.heps_ell} after splitting the
potential vorticity into a mean part $\bar{q}_\eps$ and an anomaly
$\hat q_\eps = q_\eps-\bar{q}_\eps$, and gathering all of the constant
coefficient terms on the left hand side.  In spectral space, this
leads to a simple inversion for the depth anomaly.  However, iteration
is required since $h_\eps$ appears on the right hand side multiplied
by the potential vorticity anomaly.  Nevertheless, the iteration
procedure converges rapidly in practice.  Note, we ensure that the
average anomaly is zero so that mass is exactly conserved.

Simulations are performed for a range of different $\lambda$ with a
particular focus on the cases $\lambda = 0, \tfrac12, 1$ and for a
wide range of Burger numbers (here equivalent Rossby numbers) with
$\eps=2^{-m/2}$ and $m=2,\ldots,10$.  Comparisons between the balance
and full shallow water results are made at times $t$ for
$\eps t = 0.1,\, 0.2,\, \ldots, 1$.  Differences are always evaluated on
the shallow water side by transforming the balance model fields using
the transformation detailed in Section~\ref{s.trafo}.  They are
diagnosed in the domain-averaged $L^2$-norm
\begin{align}
  \norm{\theta} 
  = \biggl( 
      \frac1{\operatorname{Vol} \Omega}
      \int_\Omega \lvert \theta \rvert^2 \, \d \x
    \biggr)^{\tfrac12} \,.
  \label{e.errordef-explicit}
\end{align}
In particular, for each of the fields $a=q$, $\delta$, and $\gamma$,
we monitor the r.m.s. difference
\begin{align}
  \mathcal{E}_a = \norm{a_\eps - \operatorname{T}[a]} \,.
  \label{e.errordef}
\end{align}

\begin{figure}
\centering
\includegraphics{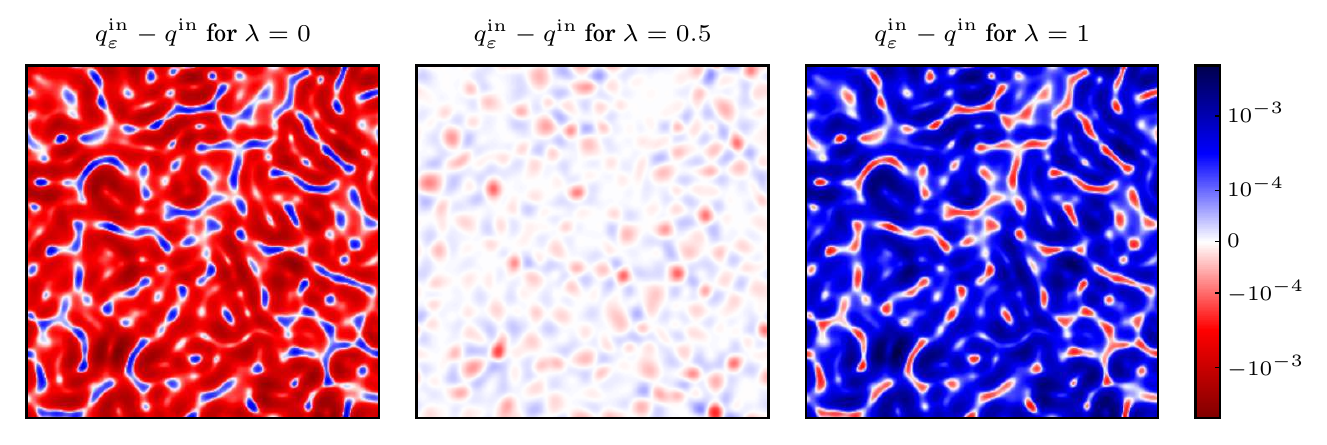}
\caption{The difference $q_\eps^\ini-q^\ini$, where
$q_\eps^\ini = T[q^\ini]$, for several values of $\lambda$ with fixed
$\eps=2^{-5}$.  Note that the color scale is logarithmic for values
above $10^{-4}$ and linear for values between $0$ and $10^{-4}$.}
\label{f.fig1}
\end{figure}

\begin{figure}
\centering
\includegraphics{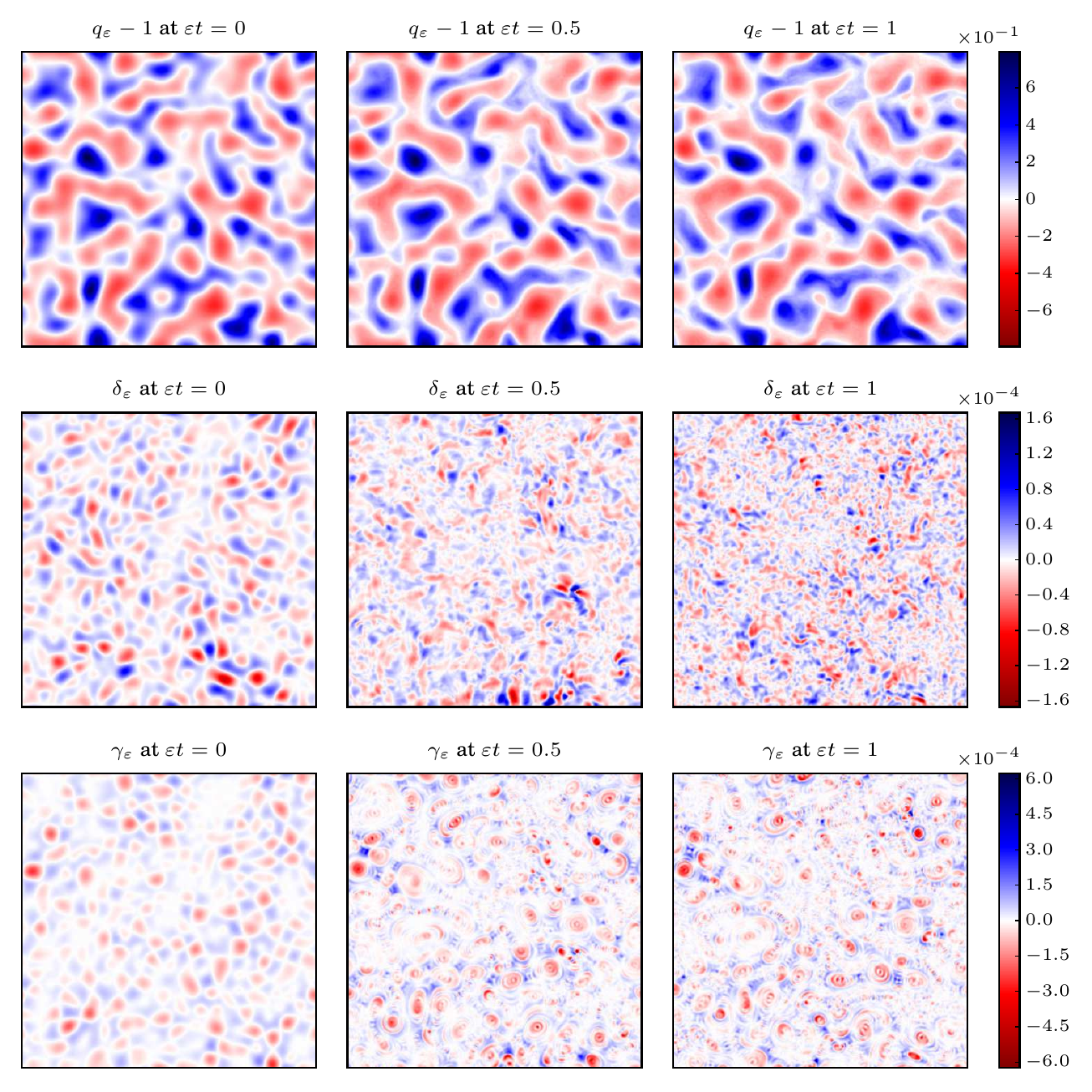}
\caption{Shallow water fields of potential vorticity anomaly
  $q_\eps-1$ (top row), divergence $\delta_\eps$ (middle row) and
  ageostrophic vorticity $\gamma_\eps$ (bottom row), for $\eps=2^{-5}$
  and initialized with $\lambda=\tfrac12$: initial time $t=0$
  (left), intermediate time $t = 0.5/\eps$ (middle), and 
  final time $t = 1/\eps$ (right).}
\label{f.fig2}
\end{figure}

\subsection{Initialization}
\label{s.initialization}

We define the characteristic horizontal length scale $L$ by the
inverse of the dominant wavenumber $k_0$, which we set to
$k_0=6$. This implies a Rossby radius of deformation of
$L_D=\sqrt{\eps}/k_0$. (Note, while a factor of $2\pi$ might seem
appropriate, $L_D$ itself is better thought of as the inverse
deformation wavenumber.) The Coriolis parameter is set to
$f=4\pi/\eps$, which then defines the gravity wave speed
$c = f \, L_D$.


The initial height $h^\ini$ on the GLSG balance model side is
generated as a random realization with a prescribed power spectrum
${\mathcal{S}}_h \sim k^3/(k^2+a\,k_0^2)^n$, taking $n=37/44$ and
$a=(2n-3)/3$ to guarantee a maximum of the spectrum at $k=k_0$.

In Figure~\ref{f.fig1}, we show the difference between the
corresponding initial potential vorticity field $q^\ini$ and the
transformed potential vorticity fields $q_\eps^\ini$, for
$\lambda=0,\tfrac12,1$ and an intermediate value of $\eps$.  Note that
this difference is not measuring the quality of the initialization or
the amount of imbalance, as $q^\ini$ and $q_\eps^\ini$ live in
different spaces.  The figure just serves to illustrate that for
$\lambda\neq \tfrac12$, the transformation produces significantly
different fields.

For $\lambda=0$ and $1$, and for $\eps=2^{-5}$, the differences between the
untransformed GLSG fields $q^\ini$ and the corresponding rotating
shallow water equation fields $q_\eps^\ini$ are about $0.06\%$,
whereas the case $\lambda=\tfrac12$ produces differences which are
$40$ times less.  This is expected as for $\lambda=\tfrac12$, by
construction, the difference between $\operatorname{T}[q^\ini]$ and
$q^\ini$ is ${\mathcal{O}}(\eps^2)$.

We remark that for $\lambda=0$ and $1$ the differences between transformed
and untransformed initial height fields are about $0.3\%$, i.e.\
almost one order of magnitude larger than the differences in the
potential vorticity fields.

\section{Results}
\label{s.numerics}
\subsection{Flow evolution}

Figure~\ref{f.fig2} demonstrates how the shallow water flow fields,
initialized by the balancing procedure described in
Section~\ref{s.trafo}, evolve on a time scale of
${\mathcal{O}}(1/\eps)$.  Shown are the potential vorticity anomaly
$q_\eps-1$, divergence $\delta_\eps$ and ageostrophic vorticity
$\gamma_\eps$ at the initial time $t=0$, an intermediate time
$t=0.5/\eps$, and the final time $t=1/\eps$, for Rossby number
$\eps=2^{-5}$. Whereas the potential vorticity appears broadly similar
at these two times, the divergence and ageostrophic vorticity exhibit
major changes.  Only a small part of these changes is due to
imbalanced motions, as seen below.

\begin{figure}
\centering
\includegraphics{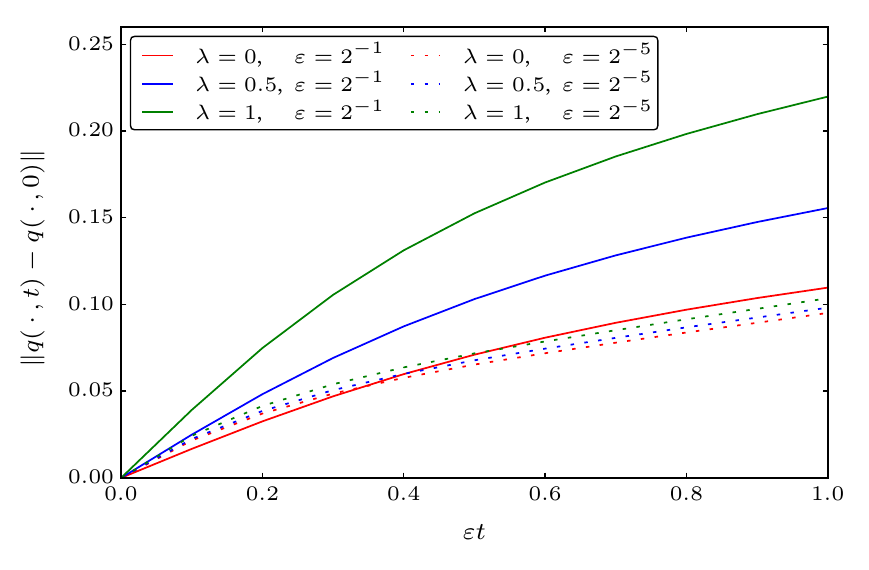}
\caption{Amount of flow evolution between $t=0$ and $t = \eps^{-1}$ as
measured by the quantity $\norm{q(\, \cdot \,,t)-q(\, \cdot \,,0)}$,
as a function of Rossby number $\eps$, for $\lambda=0$, $\tfrac12$,
and $1$, as indicated.}
\label{f.fig3}
\end{figure}

We now establish that the Eulerian evolutionary time scale is, as
theorized in Appendix~\ref{a.time-scale}, of ${\mathcal{O}}(1/\eps)$,
independent of $\lambda$. We do so by monitoring the change of the
Eulerian potential vorticity up to time $t=\eps^{-1}$.  The result is
shown in Figure~\ref{f.fig3}.  In particular, the final time
difference $\norm{q(\, \cdot \,, \eps^{-1})-q(\, \cdot \,,0)}$ is
approximately independent of the Rossby number $\eps$, and
approximately independent of $\lambda$.  This justifies evolving the
dynamics to time $t=\eps^{-1}$ to assess the order of accuracy to
which the GLSG balance models are able to approximate nearly balanced
shallow water flows.  The results of this analysis are presented below
in Section~\ref{s.asymptotic}.

\subsection{Comparison between the shallow water and GLSG dynamics}

\begin{figure}
\centering
\includegraphics{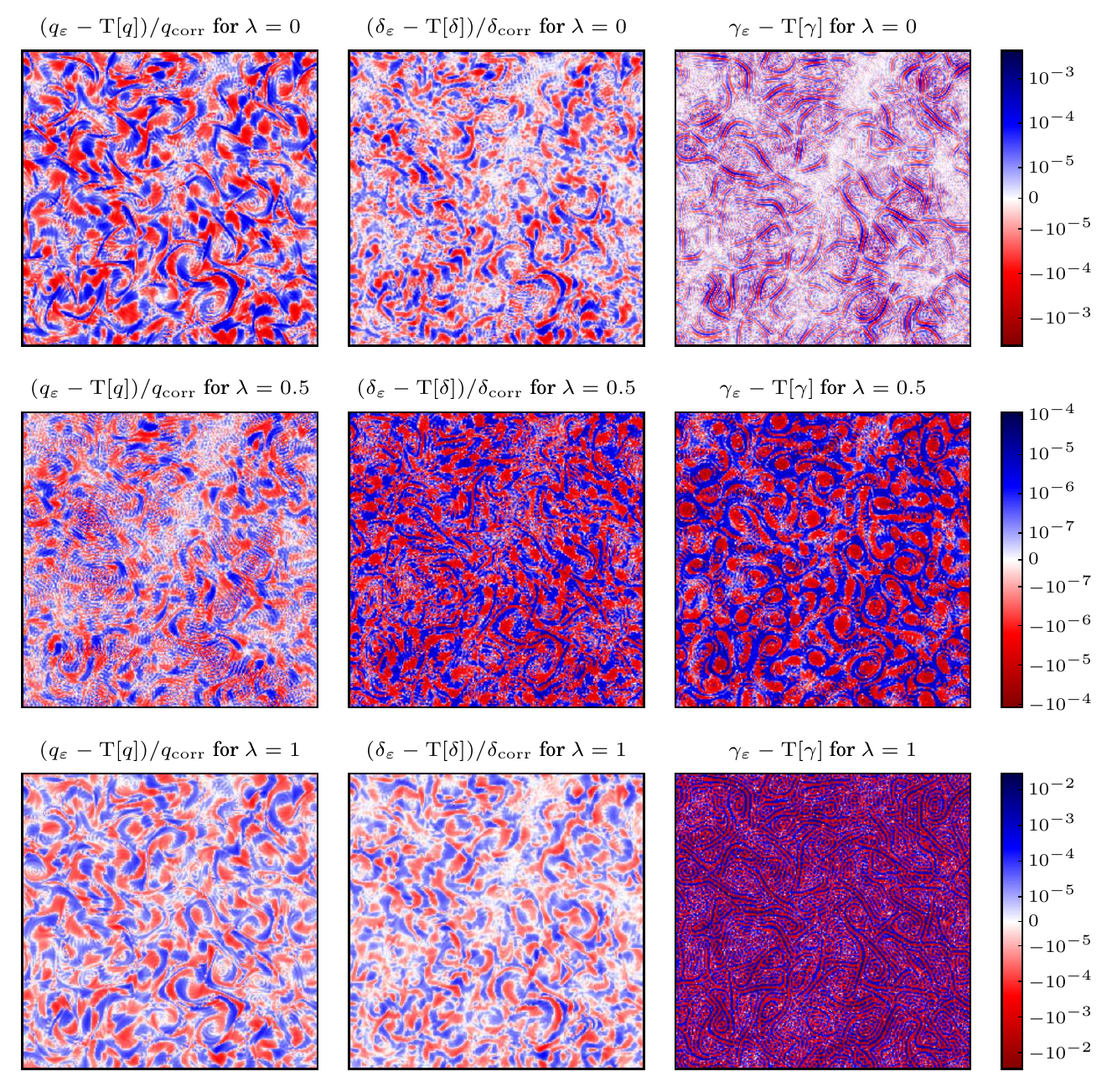}
\caption{Early time differences ($\eps t = 0.1$) between the shallow
water fields $q_\eps$, $\delta_\eps$, and $\gamma_\eps$, and the
corresponding transformed balance model fields for $\eps=2^{-5}$ and
for three different values of $\lambda$.  The correction factors
$q_{\operatorname{corr}}$ and $\delta_{\operatorname{corr}}$ are
chosen so that the differences in $q$, $\delta$, and $\gamma$ have
exactly the same range of values for the late-time ($\eps t = 1$)
frames in the most accurate case $\lambda=\tfrac12$.}
\label{f.fig4}
\end{figure}

\begin{figure}
\centering
\includegraphics{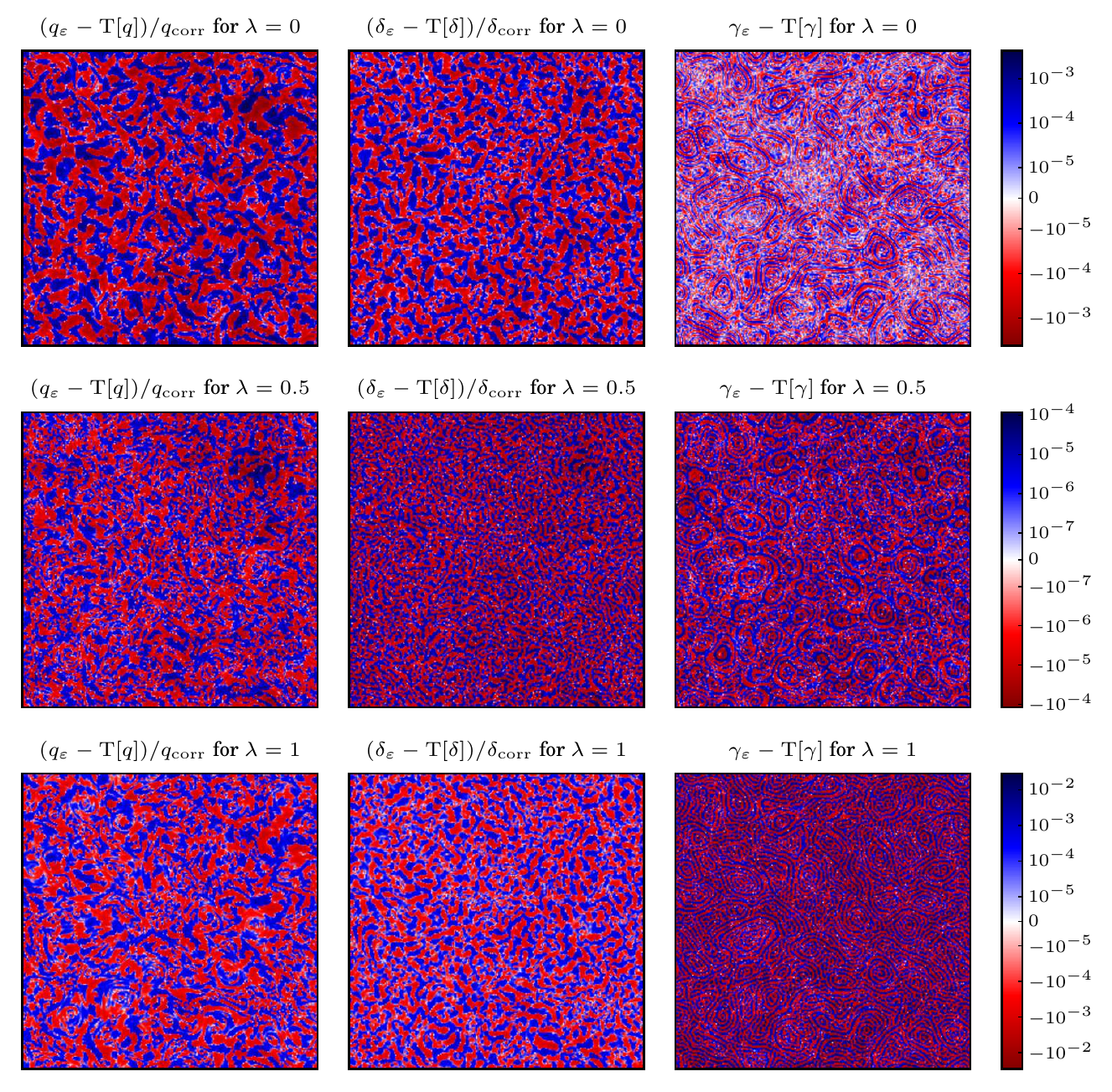}
\caption{Late time relative differences (at $\eps t=1$) of the
difference between shallow water fields and corresponding transformed
balance model fields. All other parameters and normalizing values are the same as in
Figure~\ref{f.fig4}. Note that the color
scales are the same as for the corresponding rows in
Figure~\ref{f.fig4} so that the growth in color saturation gives an
impression of the growth in the error as time progresses. }
\label{f.fig5}
\end{figure}

Before investigating the scaling behavior of the error
\eqref{e.errordef} with $\eps$, we examine the actual difference
fields between the shallow water fields $q_\eps$, $\delta_\eps$, and
$\gamma_\eps$, and the corresponding transformed balance fields
$\operatorname{T}[q]$, $\operatorname{T}[\delta]$, and
$\operatorname{T}[\gamma]$ for $\lambda=0$, $\tfrac12$, and $1$.
Note, by construction, these difference fields are identically zero at
$t=0$.  From the earliest times, we see a clear distinction between
the cases $\lambda=\tfrac12$ and $\lambda\neq \tfrac12$ --- see
Figures~\ref{f.fig4} and~\ref{f.fig5} for $t=0.1/\eps$ and $t=1/\eps$,
respectively. The differences in potential vorticity are $60$ times
smaller for $\lambda=\tfrac12$ than for $\lambda=0$ or $1$. The
differences in divergence are $15$ times smaller for
$\lambda=\tfrac12$ than for $\lambda=0$ or $1$. Both fields, however,
show similar structures. The most remarkable differences between the
cases $\lambda=\tfrac12$ and $\lambda\neq \tfrac12$ are seen in the
ageostrophic vorticity. Here the differences are $8$ times larger for
$\lambda=0$ and $250$ times larger for $\lambda=1$ when compared to
$\lambda=\tfrac12$. Moreover, whereas the structure of the difference
field resembles the actual ageostrophic vorticity field in the case
$\lambda=\tfrac12$ (cf.\ Figure~\ref{f.fig2}), the streak-like
concentration of $\gamma$ in the cases $\lambda=0$ and $\lambda=1$
appears to be unphysical.  
When $\lambda=\tfrac12$, we see ageostrophic wave-train like
structures not only in the potential vorticity difference field, but
also in the divergence and ageostrophic vorticity.  These structures
tend to be most prominent in regions of significant potential
vorticity anomalies.

\subsection{Asymptotic scaling with Rossby number}
\label{s.asymptotic}

\begin{figure}
\centering
\includegraphics{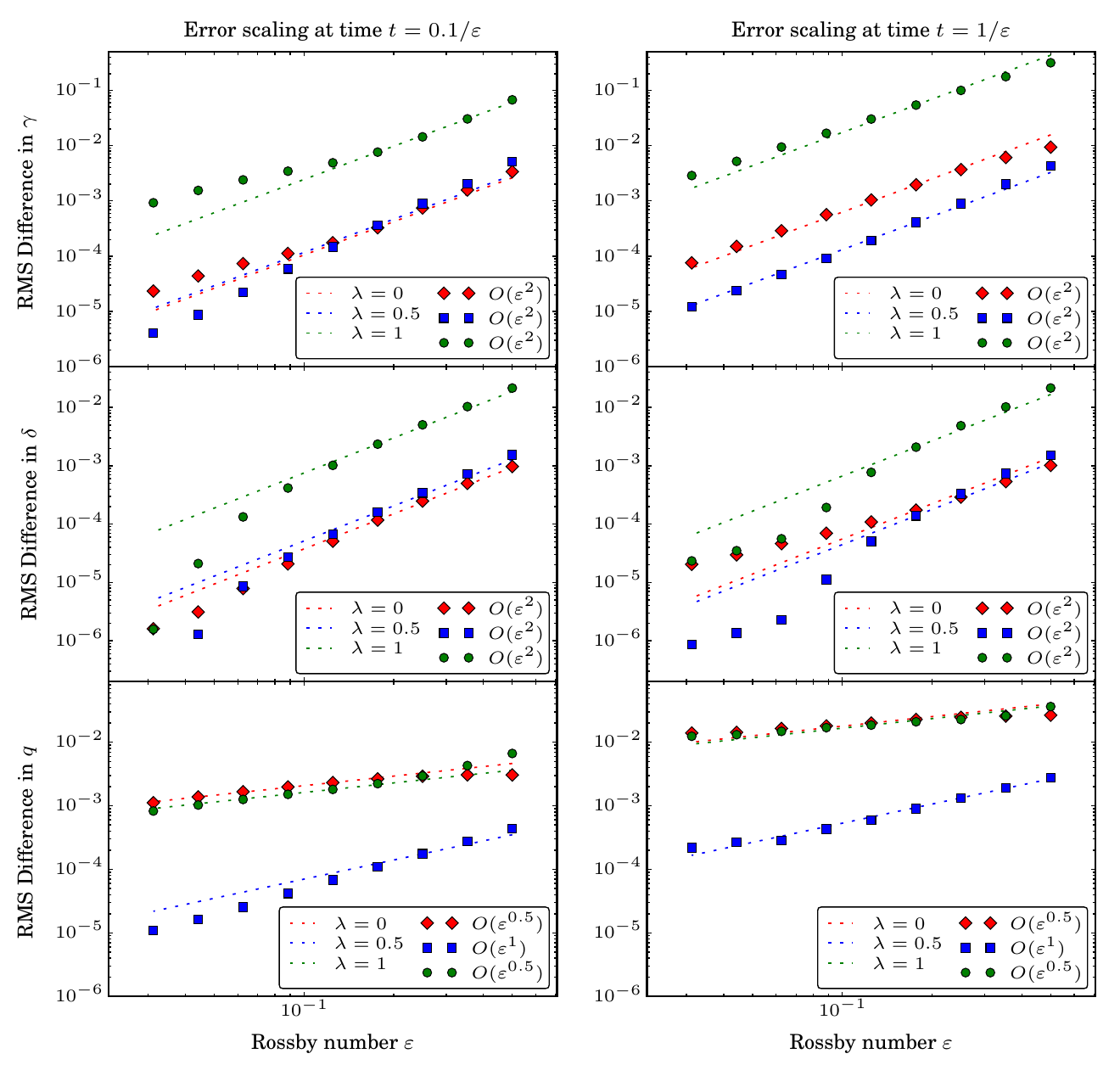}

\caption{R.m.s.\ differences $\mathcal{E}_\gamma$ (top), $\mathcal{E}_\delta$
  (middle), and $\mathcal{E}_q$ (bottom) as a function of Rossby
  number $\eps$, for various $\lambda$ as indicated, with the early
  time results ($t=0.1/\eps$) shown on the left, and the late time
  results ($t=1/\eps$) shown on the right.} 
\label{f.fig6}
\end{figure}

We next consider how the errors, as measured by the r.m.s.\
differences $\mathcal{E}_\gamma$, $\mathcal{E}_\delta$, and
$\mathcal{E}_q$, defined in \eqref{e.errordef}, scale with Rossby
number $\eps$ for various choices of $\lambda$.  These results are
presented in Figure~\ref{f.fig6}, at early time $t=0.1/\eps$ on the
left and at the final time $t=1/\eps$ on the right.  First of all, the
error grows in time, as expected, but preserves its Rossby number
scaling.  Both $\gamma$ and $\delta$ exhibit an
${\mathcal{O}}(\eps^2)$ scaling overall; the departures at small
$\eps$ are largely numerical artifacts (damping), as has been verified
in double-resolution simulations.  Most significantly, the errors in
potential vorticity (bottom panels) exhibit a shallower scaling, and
one which clearly distinguishes $\lambda=\frac12$ from
$\lambda\neq\frac12$.  Not only are the errors for
$\lambda\neq\frac12$ significantly larger than for $\lambda=\frac12$,
their scaling with $\eps$ is also significantly shallower.  This is
attributed to the poor representation of the ageostrophic dynamics for
$\lambda\neq\frac12$, already seen in Figures~\ref{f.fig4}
and~\ref{f.fig5}.

\subsection{Dependence on $\lambda$}
\label{s.lambdachoice}

\begin{figure}
\centering
\includegraphics{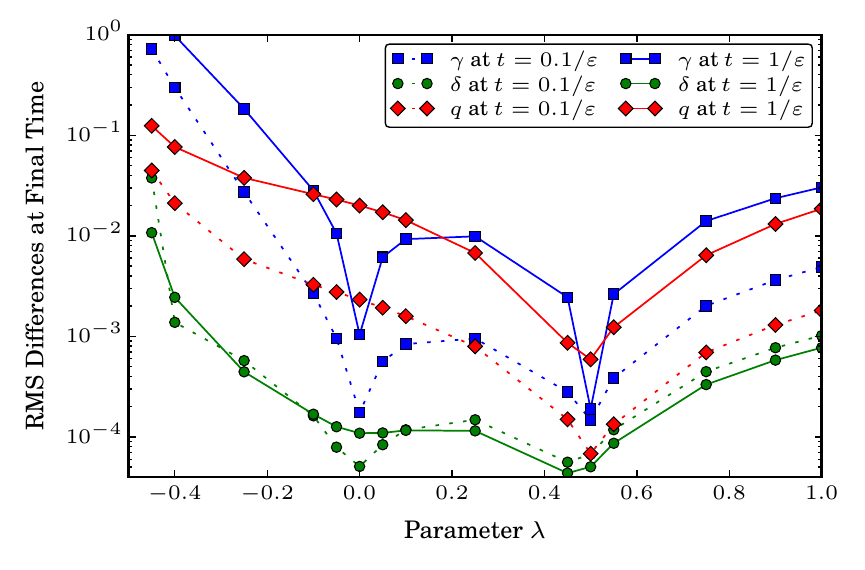}
\caption{R.m.s.\ differences $\mathcal{E}_\gamma$ (blue),
  $\mathcal{E}_\delta$ (green), and $\mathcal{E}_q$ (red) as a
  function of $\lambda$, for fixed Rossby number $\eps=2^{-3}$, at
  $t=0.1/\eps$ (dashed lines) and $t=1/\eps$ (solid lines).}
\label{f.fig7}
\end{figure}

The strikingly different behavior for different values of $\lambda$ is
seen more explicitly in Figure~\ref{f.fig7}, now showing the
dependence of the r.m.s.\ differences on $\lambda$ for a fixed Rossby
number $\eps=2^{-3}$.  Dashed lines show early time results, while the
solid lines show the final time.  There is a dip in all three error
measures at $\lambda=\frac12$, but it is most pronounced for
$\mathcal{E}_\gamma$ (blue curves).  Interestingly, a second weaker
dip occurs at $\lambda=0$, though not for $\mathcal{E}_q$.  When
$\lambda=0$ the regularity of the velocity field is expected to be
greater than that of the height field, since the high derivative terms
on the right-hand side of \eqref{e.elliptic} are absent.  This
evidently results in much smaller errors, principally in
$\mathcal{E}_\gamma$, compared to nearby surrounding values of
$\lambda$, but not as small as the errors found for $\lambda=\frac12$.
As $\lambda$ decreases further, the errors grow steeply and diverge as
$\lambda \to -\frac12$, where the balance model becomes mathematically
ill-posed.

In summary, $\lambda=\frac12$ has much weaker errors in all three
measures than any other $\lambda$, even values close to
$\lambda=\frac12$.  The exception is $\mathcal{E}_\delta$, which
appears to be less sensitive to $\lambda$.  This is consistent with
the mathematical analysis in Section \ref{ageosection}, specifically
\eqref{e.delta_eps2}, where no significant gain in regularity is seen
for $\lambda=\frac12$ (or for $\lambda=0$).  Nonetheless, even for
$\delta$, the choice $\lambda=\frac12$ leads to nearly the smallest
errors.  Most importantly, errors in potential vorticity $q$ exhibit a
single, pronounced minimum at the value $\lambda=\frac12$.  This
implies that the balance model for $\lambda=\frac12$, i.e.\ the
$L_1$-model, offers the most accurate prediction of nearly balanced
shallow water flow.

\subsection{Power spectra and regularity}
\label{s.powerspec}

\begin{figure}
\centering
\includegraphics{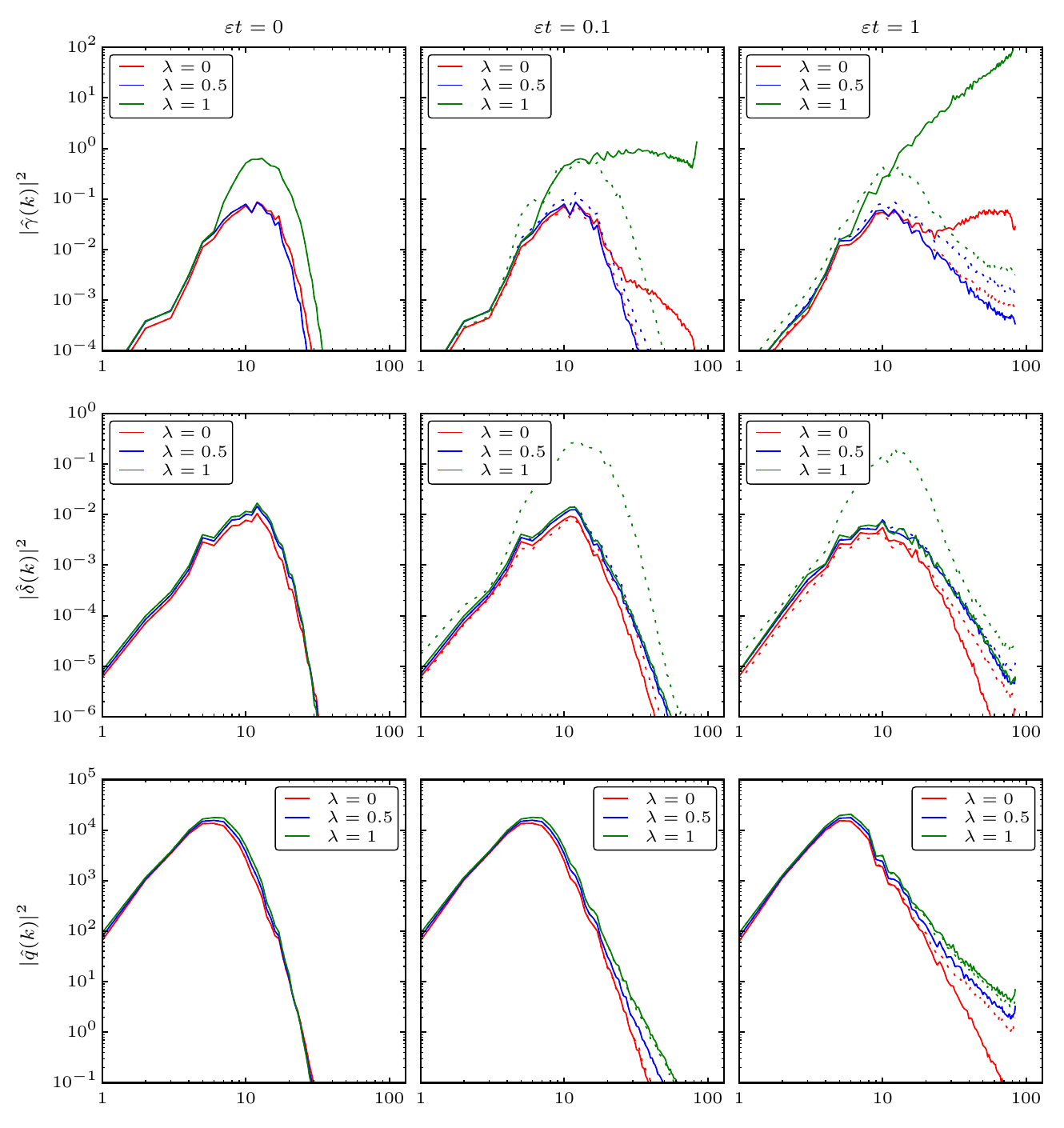}
\caption{Power density spectra for ageostrophic vorticity (top),
  divergence (middle), and potential vorticity (bottom), at times
  $t=0$ (left), $t=0.1/\eps$ (middle), and $t=1/\eps$ (right).  Here
  $\eps=2^{-3}$, and three different values of $\lambda$ are compared
  (see legend).  Dashed lines are used for the shallow water fields
  $\gamma_\eps$, $\delta_\eps$, and $q_\eps$, while solid lines are
  used for the corresponding transformed balance fields
  $\operatorname{T}[\gamma]$, $\operatorname{T}[\delta]$, and
  $\operatorname{T}[q]$.}
\label{f.fig8}
\end{figure}

In Figure~\ref{f.fig8} we show power spectra for potential vorticity,
divergence and ageostrophic vorticity, both on the GLSG balance model
side and on the shallow water side. Whereas the spectra of potential
vorticity and of divergence each exhibit closely similar forms for the
different values of $\lambda$ and model dynamics, the ageostrophic
vorticity spectra exhibit large differences from the earliest times.
The ageostrophic vorticity spectra for $\lambda=0$ and $\lambda=1$
rapidly develop strong high-wavenumber contributions which dominate
the spectra. This erroneous behavior corresponds to the intense
frontal structures seen in the difference fields in
Figures~\ref{f.fig4} and~\ref{f.fig5}. By contrast, the ageostrophic
vorticity spectrum for $\lambda=\tfrac12$ exhibits a closely similar,
decaying form on both the GLSG and shallow water sides at all times.

It is also noteworthy that, only for $\lambda=\tfrac12$, the spectrum
for $\operatorname{T}[q]$ is steeper than that of the corresponding
$q_\eps$. This is to be expected for a reliable balance model, as it
allows for higher wavenumber contributions of inertia-gravity waves in
the shallow water equations expressing the departure from balance.

\begin{figure}
\centering
\includegraphics{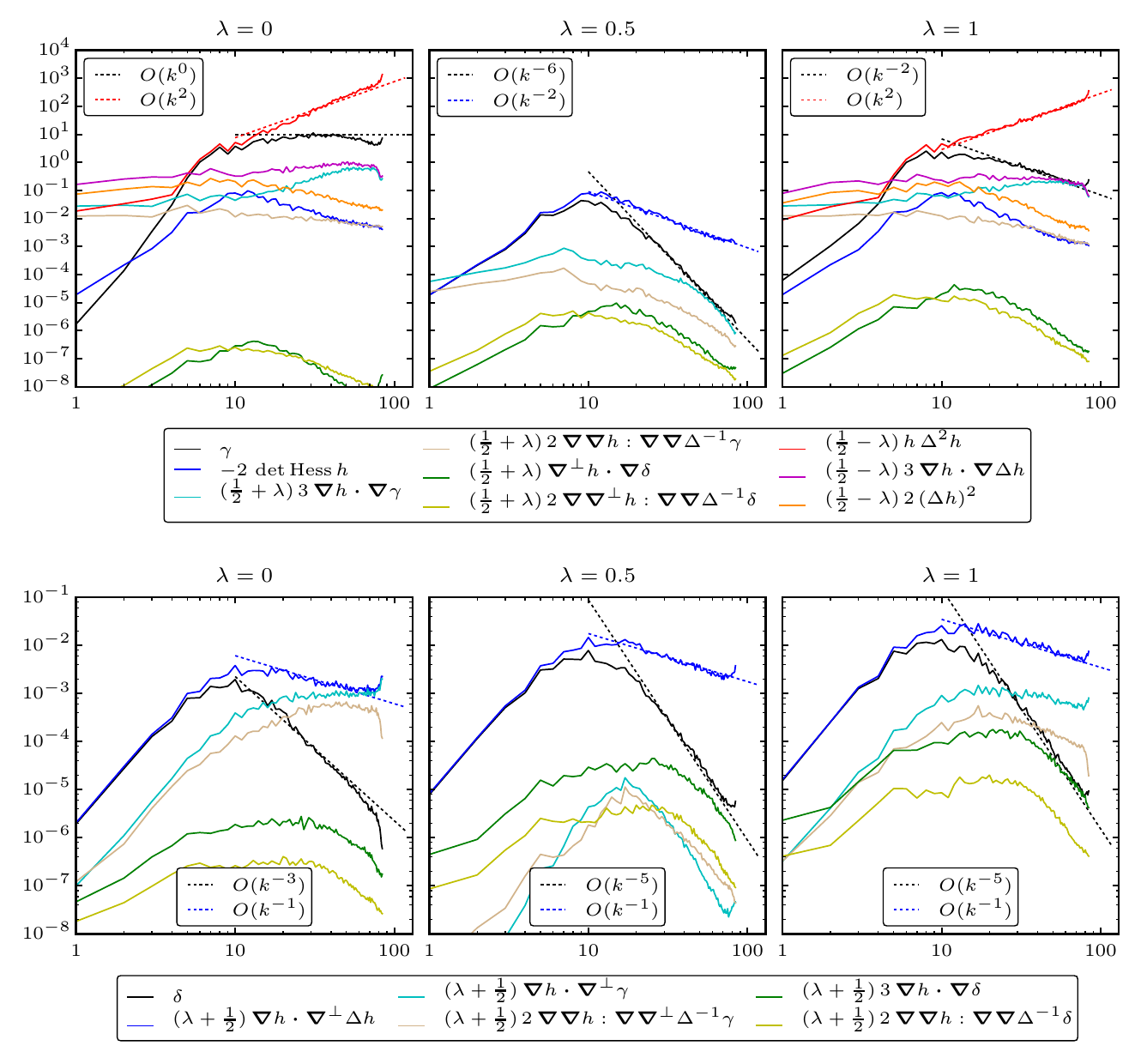}
\caption{Power density spectra for the balance model $q$, $h$, and
$\u$  for
$\eps=2^{-3}$, and for different values of $\lambda$ at time $\eps t
= 1$. }
\label{f.fig10}
\end{figure}

\begin{figure}
\centering
\includegraphics{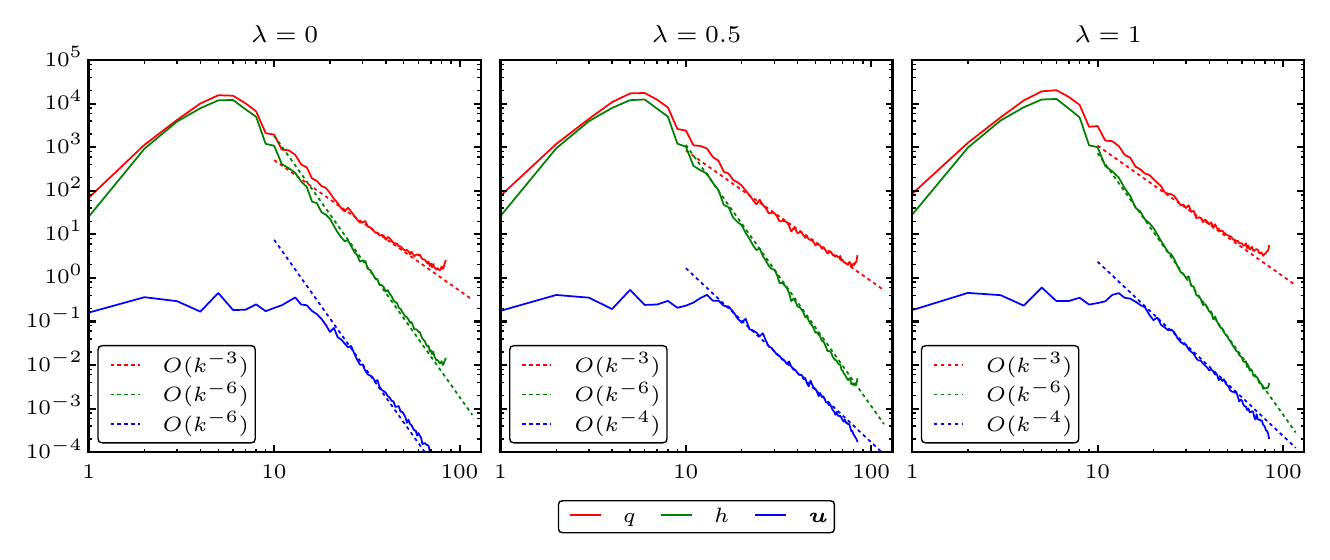}
\caption{Power density spectra for each of the terms on the right hand
side of the $\gamma$-equation \eqref{e.ageovort_eps2}, upper panel,
and of the $\delta$-equation \eqref{e.delta_eps2}, lower panel,
corresponding to the case shown in Figure~\ref{f.fig10}.}
\label{f.fig9}
\end{figure}

We now look at the terms affecting the regularity of the solutions to
the GLSG balance relation in more detail.  To help interpret the results
properly, we note that when the Fourier coefficients $a_\k$ of a
two-dimensional field $a$ decay like
$\lvert \k \rvert^{-p} \equiv k^{-p}$, then the power spectrum decays
like $k^{1-2p}$.

Figure~\ref{f.fig10} shows the power spectra of $q$, $h$, and $\u$.
The power spectrum of $h$ decays robustly like $k^{-6}$ independent of
$\lambda$, corresponding to $h_\k \sim k^{-7/2}$. The regularity of
$\u$ is best when $\lambda=0$. However, since for $\lambda=0$ the
balance relation \eqref{e.elliptic} implies that $\u$ is one
derivative smoother than $h$, we would expect a decaying velocity
power spectrum proportional to $k^{-8}$. The observed reduced spectral
decay is presumably due to nonlinear effects.  The lack of regularity
for $\lambda \neq \tfrac12$ noted above predominantly affects the
\emph{ageostrophic} part of the flow, whereas $q$, $h$, and $\u$ are
dominated by the geostrophic part of the flow which masks the
deterioration of the smaller ageostrophic part.

Diagnosing the balance relation in ageostrophic variables, however,
offers an explanation for the observed deterioration when
$\lambda \neq \tfrac12$.  This is done in Figure~\ref{f.fig9}, which
displays the final time power density spectra for each of the terms on
the right hand side of the $\gamma$-equation \eqref{e.ageovort_eps2}
and of the $\delta$-equation \eqref{e.delta_eps2}.  The term with the
highest number of derivatives on the right hand side of the equation
for $\gamma$, namely $(\tfrac12-\lambda) \, h \, \Delta^2h$, gives
rise to a spectrum \emph{increasing} as $k^2$ for both $\lambda=0$ and
$\lambda=1$.  As a result, even though the elliptic operator on the
left hand side gains some regularity, the spectrum of $\gamma$ shows
no decrease when $\lambda=0$, and only a slight decrease when
$\lambda=1$.  This saturation at high wavenumbers is unphysical.
When $\lambda=\tfrac12$, the dominant term on the right hand side of
\eqref{e.ageovort_eps2} has a power spectrum decaying like $k^{-2}$
and the elliptic inversion gains the expected two derivatives, so that
the power spectrum of $\gamma$ decays like $k^{-6}$.

The equation for $\delta$, \eqref{e.delta_eps2}, does not have any
$\lambda$-dependent irregular terms on its right hand side and is
therefore much less sensitive to $\lambda$.  However, it is clearly evident
that when $\lambda \neq \tfrac12$, the poor spectral decay of $\gamma$
contaminates some of the normally sub-dominant terms on the right hand side
to reduce the spectral decay of $\delta$.  This is particularly evident
when $\lambda=0$.

For $\lambda=\tfrac12$ in particular, the
right hand side of \eqref{e.ageovort_eps2} is dominated by two
derivatives on $h$.  Since $h_\k \sim k^{-7/2}$ as observed in 
Figure~\ref{f.fig10}, then the right hand side of \eqref{e.ageovort_eps2}
should exhibit a Fourier decay like $k^{-3/2}$,
resulting in $\gamma_\k \sim \lvert k \rvert^{-7/2}$.  The
corresponding dominant term on the right hand side of
\eqref{e.delta_eps2} for $\delta$, namely
$\grad h \Cdot \grad^\bot \Delta h$, contains three derivatives on
$h$, and thus is expected to have a flat power spectrum.  Yet, the
observed power spectrum for this term decays like $k^{-1}$,
corresponding to a $k^{-1}$ decay of its Fourier coefficients, and
therefore $\delta_\k \sim k^{-3}$ after inversion of the elliptic
operator.  Its spectrum is steeper than the spectrum of the term
$\grad h \Cdot \grad \Delta h$ on the right hand side of
\eqref{e.ageovort_eps2}, which suggests that there is some
cancellation within the nonlinear contributions that is not currently
understood.

We finally remark that the absolute slopes see in Figures~\ref{f.fig10}
and~\ref{f.fig9} do not represent a late-time steady state
characterized by sharp potential vorticity gradients.  At this stage
of the evolution, they are still in the process of steepening.  The
relative slopes, however, are robust.


\section{Discussion and outlook}
\label{s.summary}

We have examined a family of variational balance models relevant to
the small Rossby number, semi-geostrophic regime of the rotating
shallow water equations.  This family, originally derived by
\citet{Oliver06}, is spanned by a parameter $\lambda$ and includes the
$L_1$-model introduced by \citet{Salmon83} as a special case
($\lambda=\frac12$).  To test the quality of these models, we have
compared them against initially balanced shallow water numerical
simulations for a wide range of Rossby numbers $\eps$.  This has
revealed that the $L_1$-model, obtained for the specific parameter
value $\lambda=\frac12$, strongly outperforms all other members of the
family.  That is, the $L_1$-model gives the closest comparison with
the full shallow water dynamics over an ${\mathcal{O}}(\eps^{-1})$
time scale.  Given that all models are formally of the same asymptotic
order, and that the case $\lambda=0$ seems preferable from the
regularity theory point of view, this result was initially unexpected.
However, we have been able to explain the superior performance of the
$L_1$-model by rewriting the balance model in terms of ageostrophic
quantities, where the ageostrophic vorticity is most regular when
$\lambda=\tfrac12$.  Our numerical diagnostics confirm that this
interpretation is consistent with the actual model behavior.  In
particular, the ageostrophic vorticity for $\lambda=\frac12$ exhibits
a steeply decaying spectrum, in close agreement with the full shallow
water dynamics.  On the other hand, the ageostrophic vorticity for
$\lambda\neq\frac12$ exhibits a flat or rising spectrum.  This
unphysical feature spoils the comparison with the full shallow water
dynamics.  This finding underscores the critical importance of
understanding, and ensuring, the mathematical regularity of any
balance model.  We further remark that the observed superior
performance of the $L_1$ model is consistent with the study of
\cite{AllenHN2002TowardEG} who find that the stratified version of the
$L_1$ model and its next order correction outperform a selection of
other balance models in a simple direct numerical comparison.

Over longer time scales, randomly initialised shallow-water flows
generically exhibit a direct enstrophy (potential vorticity variance)
cascade to small scales, leading to sharp fronts and fine scale
filamentary debris, particularly in potential vorticity.  As a result,
predictability is first lost at small scales then progressively at
larger scales due to nonlinear scale interactions.  This makes any
direct comparison with a balance model difficult, though it may still
be meaningful to compare statistical properties.  The methods used in
the present paper were designed to address how different balance
models compare to the full shallow water model \emph{before} any
significant small-scale structure develops. i.e.\ while the flow is
still predictable at all scales considered.  Different methods, better
suited to preserving conservation laws (to the extent possible), would
be needed to study both the balanced and shallow-water dynamics at
longer times, e.g.\ as in \cite{MohebalhojehDritschel01}.

There are several new ideas to pursue emerging from the work presented
here.  We have focused above on a particular form of the initial
conditions. It would be interesting to see how the balanced GLSG
models perform in flows starting from a few, well-separated vortical
structures, and where the largest velocity gradients are concentrated
in thin jets of width comparable to the Rossby radius of deformation
$L_D$. Notably, the balance relation \eqref{e.elliptic} exhibits
consistent scaling in this scenario.  The concentration of fluid flow
in jets of width $L_D$ implies $u\sim f L_D\sim \sqrt{\eps}$ and
$\grad\sim 1/\sqrt{\eps}$. Since $\u\approx \grad^\bot h$ we have
$h\sim \eps$. Assuming that the jets are characterized (in the worst
case) by jumps in potential vorticity, which have at worst a spectral
scaling $q_\k \sim k^{-1}$, we have $\u_\k \sim k^{-2}$ and
$h_\k \sim k^{-3}$. The balance relation \eqref{e.elliptic} is
invariant under this scaling.

In future work, we plan to compare the GLSG balance models studied
here with other models used in the literature.  In particular, it will
be instructive to see how the geometric GLSG equations compare with
more traditional balance models obtained by performing the asymptotics
directly to the equations of motion.  Examples include the
$\delta$-$\gamma$ hierarchy of balance models introduced by
\citet{MohebalhojehDritschel01} and the semi-geostrophic equations
which are presumed valid specifically in the frontal regime
\citep{cullen2008comparison}. 

Of particular interest is the behavior of the $L_1$-model, and
possibly other models from the GLSG family, in spherical geometry.  At
the formal level, the variational derivation of the models should
translate naturally to spherical geometry.  However, it is less clear
whether the resulting balance models remain mathematically well posed
and can be simulated in a robust way as the Coriolis parameter
degenerates at the equator.  Previous work by \cite{Oliver:2013}
suggests that robust solvability at mid-latitudes may only be possible
if the transformation vector field $\v$ is nontrivial at leading
order, i.e., if one moves away from Salmon's $L_1$ model.  This work
would need to be revisited in the light of rewriting the balance
relation in terms of ageostrophic quantities.  An independent issue is
the study of degeneracy near the equator.  We plan to address these
questions in future work.


Although we have not considered the problem of quantifying the amount
of imbalance (or gravity wave activity) associated with the
initialization procedure in this work, our framework permits us to do
so. By employing the dynamic global iteration rebalancing procedure,
described in Appendix~\ref{a.dynbal}, we can compute at each time step
the difference between the time-evolved shallow water fields and their
rebalanced forms. If the balance model used to rebalance the fields
is accurate, the difference would be dominated by gravity waves, at
least for small Rossby numbers. This could be tested by looking at the
frequency spectra of those rebalanced differences. This is planned for
future work.


\appendix
\section{Derivation of the balance model Euler--Lagrange equation}
\label{a.oliver}
Let us now compute the variation of each of the four terms appearing
in $L_\bal$ in \eqref{e.l-bal}.  First, up to perfect time derivatives which are
null-Lagrangians as they do not contribute to the variation of the
action integral,
\begin{align}
  \delta \int & \R \circ \Eta \Cdot \dot \Eta \, \d\a
    = \int \bigl[
             \grad \R \circ \Eta \, \delta \Eta \Cdot \dot \Eta
             + \R \circ \Eta \Cdot \delta \dot \Eta
           \bigr] \, \d\a
      \notag \\
  & = \int \bigl[
             \grad \R^T \circ \Eta \, \dot \Eta \Cdot \delta \Eta 
             - \grad \R \circ \Eta \, \dot \Eta \Cdot \delta \Eta
           \bigr] \, \d\a
    = - \int h \, \w \Cdot \u^\bot \, \d\x \,.
  \label{e.variation1}
\end{align}
The last identity holds as $\grad^\bot \Cdot \R = 1$, which implies
that $\grad \R - \grad \R^T = \mathsf J$, the standard symplectic matrix.

Second,
\begin{align}
  \delta \int & \grad^\bot h \circ \Eta \Cdot \dot \Eta \, \d\a
    = \int \bigl[
             \grad^\bot \delta h \circ \Eta \Cdot \dot \Eta
             + \grad \grad^\bot h \circ \Eta \, \delta \Eta 
               \Cdot \dot \Eta
             + \grad^\bot h \circ \Eta \Cdot \delta \dot \Eta
           \bigr] \, \d\a
      \notag \\
  & = \int \bigl[
             \grad^\bot \delta h \circ \Eta \Cdot \dot \Eta
             + \grad^\bot \grad h \circ \Eta \, \dot \Eta 
               \Cdot \delta \Eta
             - \grad^\bot \dot h \circ \Eta \Cdot \delta \Eta
             - \grad \grad^\bot h \circ \Eta \, \dot \Eta
               \Cdot \delta \Eta
           \bigr] \, \d\a
      \notag \\
  & = \int h \, \bigl[
             - \grad^\bot \grad \Cdot (h\w) \Cdot \u
             + \grad^\bot \grad h \, \u \Cdot \w
             + \grad^\bot \grad \Cdot (h\u) \Cdot \w
             - \grad \grad^\bot h \, \u \Cdot \w
           \bigr] \, \d\x
      \notag \\
  & = \int h \, \bigl[
             - \grad \grad^\bot \Cdot (h\u) \Cdot \w
             + \grad^\bot \grad h \, \u \Cdot \w
             + \grad^\bot \grad \Cdot (h\u) \Cdot \w
             - \grad \grad^\bot h \, \u \Cdot \w
           \bigr] \, \d\x
      \notag \\  
  & = \int h \, \bigl[ 
             h \, \Delta \u^\bot + 2 \, \grad h \Cdot \grad \u^\bot
           \bigr] \Cdot \w \, \d\x \,,
\end{align}
again up to perfect time derivatives, which we have subtracted in the
second equality (equivalent to integration by parts with respect to
time under the action integral).  In the third equality, we have
changed to Eulerian variables and have made use of the momentum
and continuity equations.  The fourth equality results from
an integration by parts, and the last equality is straightforward
vector algebra.

Third, 
\begin{align}
  \frac12 \, \delta \int h^2 \, \d\x
  = \int h \, \delta h \, \d \x
  = - \int h \, \grad \Cdot (h \w) \, \d \x
  = \int h \, \w \Cdot \grad h \, \d\x \,.
\end{align}

Fourth,
\begin{align}
  \delta \int h \, \lvert \grad h \rvert^2 \, \d\x
  & = \int \bigl[
             \delta h \, \lvert \grad h \rvert^2
             + 2 \, h \, \grad h \Cdot \grad \delta h 
           \bigr] \, \d\x
      \notag \\
  & = - \int \bigl[
             \grad \Cdot (h\w) \, \lvert \grad h \rvert^2
             + 2 \, h \, \grad h \Cdot \grad \grad \Cdot (h\w)
           \bigr] \, \d\x
      \notag \\
  & = \int h \, \w \Cdot \bigl[
             \grad \lvert \grad h \rvert^2
             - 2 \, \grad \grad \Cdot (h \grad h)
           \bigr] \, \d\x
      \notag \\
  & = - \int h \, \w \Cdot \grad \bigl[
             2 \, h \, \Delta h
             + \lvert \grad h \rvert^2
           \bigr] \, \d\x \,.
  \label{e.variation4}
\end{align}
Plugging the results from \eqref{e.variation1} to \eqref{e.variation4}
back into the variation of the action associated with \eqref{e.l-bal},
we find that stationary points of this action imply the
Euler--Lagrange equation \eqref{e.elliptic}.


\section{Time scale of the Eulerian dynamics}
\label{a.time-scale}

To leading order, the motion induced by a velocity field computed from
\eqref{e.elliptic} is geostrophic with an ${\mathcal{O}}(1)$ velocity.
Thus, fluid parcels travel a unit distance over times of
${\mathcal{O}}(1)$.  The question is: on what time scale do Eulerian
quantities change?  To answer this, we conduct a kinematic analysis,
in which we assume that $\u$ is constrained by the balance relation
\eqref{e.elliptic}, and then estimate the magnitude of $\partial_t h$
and $\partial_t h_\eps$.

First, we rearrange the balance relation \eqref{e.elliptic} so that
\begin{equation}
  \u^\bot + \, \grad h
  = \eps \, 
    \bigl[
      (\lambda + \frac12) \, (h \, \Delta \u^\bot
        + 2 \grad h \Cdot \grad \u^\bot)
      + \lambda \, \grad (2 \, h \, \Delta h
        + \lvert \grad h \rvert^2)
    \bigr] \,.
  \label{e.elliptic-1}
\end{equation}
Re-insertion of leading-order geostrophic balance into
\eqref{e.elliptic-1} gives $\u = \grad^\bot h - \eps \, \w^\bot +
{\mathcal{O}}(\eps^2)$ with 
\begin{align}
  \w = (\lambda - \frac12) \, h \, \Delta \grad h
       -  \grad h \Cdot \grad \grad h 
       + 2 \lambda \, \grad h \, \Delta h \,.
  \label{e.elliptic-2}
\end{align}
Inserting $\u = \grad^\bot h - \eps \, \w^\bot +
{\mathcal{O}}(\eps^2)$ into the transformation \eqref{e.v}, we obtain
\begin{equation}
  \v = (\lambda - \frac12) \, \grad h + \tfrac12 \, \eps \, \w 
  + {\mathcal{O}}(\eps^2) \,.
\end{equation}
Similarly, inserting \eqref{e.elliptic-2} into the continuity
equation \eqref{e.cont} gives
\begin{align}
  \dot h 
  & = - \grad \Cdot (h \u) 
    = \eps \, \grad \Cdot (h \w^\bot) + {\mathcal{O}}(\eps^2)
      \notag \\
  & = \eps  \, 
      \bigl[
        h \, \grad^\bot h \Cdot \grad \Delta h
        + \grad^\bot h \Cdot \grad \grad h \, \grad h 
      \bigr] + {\mathcal{O}}(\eps^2) \,.
  \label{e.doth2}
\end{align}
This shows that the time scale of Eulerian evolution in balance model
coordinates is ${\mathcal{O}}(1/\eps)$ and is, in particular,
independent of $\lambda$ at this order.  Moreover, we see that the
time derivative of the transformation vector field vanishes to
$\mathcal O(\eps)$:
\begin{equation}
  \dot \v = (\lambda - \frac12) \, \grad \dot h + {\mathcal{O}}(\eps)
          = {\mathcal{O}}(\eps) \,.
\end{equation}

A similar computation can be performed after transforming to the
shallow water side.  Using the diagnostic expressions for $h'$ and
$\u'$, equations \eqref{e.hp} and \eqref{e.up}, respectively, we
compute
\begin{align}
  \partial_t h_\eps
  & = - \grad \Cdot (h_\eps \u_\eps) 
    = - \grad \Cdot (h \u)
      - \eps \, \grad \Cdot (h' \u + h \u')
      + \mathcal O (\eps^2) \notag \\
  & = -\grad \Cdot (h \grad^\bot h - \eps \, h \, \w^\bot)
      \notag \\
  & \quad
      + \eps \,  (\lambda - \frac12) \, \grad \Cdot
        \bigl(
          \grad \Cdot (h \grad h) \, \grad^\bot h
          + h \, (\grad^\bot \grad^\bot h - \grad \grad h) \, 
            \grad^\bot h
        \bigr) + \mathcal O (\eps^2) \notag \\
  & = \tfrac14 \, \eps  \, 
        \bigl(
          \grad^\bot h \Cdot \grad \Delta h^2
          - \grad^\bot \Delta h \Cdot \grad h^2  
        \bigr) + \mathcal O (\eps^2) \,.
\end{align}
Thus, we obtain the same conclusion in shallow water coordinates as
expected by consistency of the asymptotic derivation.  In particular,
the leading ${\mathcal{O}}(\eps)$-term is independent of $\lambda$.


\section{The inverse transformation}
\label{a.dynbal}

In our setting, the transformation from balance model coordinates to
physical coordinates is explicit and has been detailed in
Section~\ref{s.trafo}.  However, it is also possible to invert the
transformation in the following sense: given a shallow water potential
vorticity $q_\eps$ in physical coordinates, we seek a corresponding
height field $h_\eps$ and velocity field $\u_\eps$, also in physical
coordinates (or, equivalently, the divergence $\delta_\eps$,
ageostrophic vorticity $\gamma_\eps$, and velocity mean
$\bar \u_\eps$) which, on the one hand, are consistent with the
definition of the shallow water potential vorticity,
\begin{equation}
  \label{e.ertl-pv}
  q_\eps = \frac{1 + \eps \, \grad^\bot \Cdot \u_\eps}{h_\eps} \,,
\end{equation}
and, on the other hand, are consistent with the balance relation
\eqref{e.elliptic} in transformed variables under the transformation
\eqref{e.trafo}.  This can be achieved as follows.

We start by decomposing $q_\eps = \bar q_\eps + \hat q_\eps$, where
$\bar q_\eps$ denotes the mean value of $q_\eps$ and $\hat q_\eps$
denotes the deviation from the mean, with corresponding notation for
the other field variables. The expression for potential vorticity
\eqref{e.ertl-pv} can then be written in the form
\begin{equation}
  (\bar q_\eps - \eps \Delta) \hat h_\eps
  = 1 - \bar q_\eps 
    - \hat q_\eps \, h_\eps + \eps \, \gamma_\eps \,, 
  \label{e.pviter2}
\end{equation}
where we have used the definition of ageostrophic vorticity
$\gamma_\eps=\grad^\bot \Cdot \u_\eps - \Delta h_\eps$, as well as the
fact that the mean height $\bar h_\eps=1$.  Equation \eqref{e.pviter2}
can be solved by iteration provided $\hat q_\eps$ is sufficiently
small. Next, to determine consistent balanced GLSG fields, we
interpret the transformation of potential vorticity in the
Lagrangian variables as
\begin{equation}
  q_\eps \circ \Eta_\eps = q \,,
\end{equation}
which leads to an advection equation with $\eps$ playing the role of
time, namely
\begin{equation}
  q_\eps' + \v_\eps \Cdot \grad q_\eps = 0 \,,
  \label{e.q-advection}
\end{equation}
where the prime denotes differentiation with respect to $\eps$ and we
integrate backwards from the given value of $\eps$ to $\eps=0$.  Of
course, we cannot have knowledge of the full transformation vector
field $\v_\eps$ as that would be akin to having an all-order balance
model.  For a first order model, it is consistent to approximate
$\v_\eps$ by $\v$ as given by \eqref{e.v}.  Thus, numerically, we are
solving
\begin{gather}
  q_\eps' + \v \Cdot \grad q_\eps = 0 
  \label{e.q-advection-approx}
\end{gather}
as a backward advection equation with $\eps$ as the artificial time
variable. 

The global iteration loop is then as follows. Given an initial
potential vorticity field $q_\eps$ on the shallow water side,
initialize the iteration with $\u_\eps^\ag=0$ (implying
$\gamma_\eps=0$ and $\delta_\eps=0$).  On the balance model side,
initialize $q = q_\eps$ and find initial $h$, $\u$, and $\v$ as in
Steps~3--5 below.
\begin{enumerate}[{Step}~1:\;\;\;]
\item Compute the corresponding height field $h_\eps$ using
\eqref{e.pviter2}.
\item Compute the potential vorticity $q$ on the balanced GLSG side by
backwards advection in $\eps$ to $\eps=0$ using
\eqref{e.q-advection-approx}.
\item Compute the balanced GLSG height field $h$ via potential vorticity
inversion \eqref{e.sg-pv-inv}.
\item Compute the corresponding GSLG velocity field $\u$ using the balance
relation \eqref{e.elliptic}.
\item Compute $\v$ via \eqref{e.v}.
\item Transform back to the shallow water side using \eqref{e.trafo}
to compute $h_\eps$ and $\u_\eps$.
\item Update the ageostrophic velocity $\u_\eps^\ag$, and go to Step 1.\\
\end{enumerate}
Repeat until a fixed point is reached.  Empirically, the procedure
converges for small to moderate values for $\eps$, but may fail to
converge when $\eps \approx 1$.

The procedure outlined above allows one to ``rebalance'' a given state of
the shallow water evolution using any of the GLSG balance models.
Given only the potential vorticity, all other fields can be reconstructed
consistent with the balance relation, and the residual can be taken as
a measure of imbalance.

\section*{Acknowledgments}

We thank Mike Cullen, Darryl Holm, and Sergiy Vasylkevych for
interesting discussions on the benchmarking of balance models.  MO's
initial work on this subject was supported by German Science
Foundation grant OL-155/3.  Further, this paper contributes to the
project ``Systematic Multi-Scale Modeling and Analysis for Geophysical
Flow'' of the Collaborative Research Center TRR 181 ``Energy Transfers
in Atmosphere and Ocean'' funded by the German Research Foundation.
Funding through the TRR 181 is gratefully acknowledged.  GAG's initial work was funded by the Australian Research Council grant DP0452147. All three
authors received support for this research from the UK Engineering and
Physical Sciences Research Council (grant number EP/H001794/1).



\begin{thebibliography}{31}
\expandafter\ifx\csname natexlab\endcsname\relax\def\natexlab#1{#1}\fi

\bibitem[Allen \& Holm(1996)]{allen1996extended}
{\sc Allen, J.~S. \& Holm, D.~D.} 1996 Extended-geostrophic {H}amiltonian
  models for rotating shallow water motion. {\em Phys. D\/} {\bf 98}~(2),
  229--248.

\bibitem[Allen {\em et~al.\/}(2002)Allen, Holm \&
  Newberger]{AllenHN2002TowardEG}
{\sc Allen, J.~S., Holm, D.~D. \& Newberger, P.~A.} 2002 Toward an
  extended-geostrophic {E}uler--{P}oincar\'e model for mesoscale oceanographic
  flow. In {\em Large-scale atmosphere--ocean dynamics\/} (ed. J.~Norbury \&
  I.~Roulstone), vol.~1, pp. 101--125. Cambridge University Press.

\bibitem[Benamou \& Brenier(1998)]{benamou1998weak}
{\sc Benamou, J.~D. \& Brenier, Y.} 1998 Weak existence for the semigeostrophic
  equations formulated as a coupled {M}onge--{A}mp\`ere/transport problem. {\em
  SIAM J. Appl. Math.\/} {\bf 58}~(5), 1450--1461.

\bibitem[Bloom {\em et~al.\/}(1996)Bloom, Takacs, Da~Silva \&
  Ledvina]{BloomEtAl96}
{\sc Bloom, S.~C., Takacs, L.~L., Da~Silva, A.~M. \& Ledvina, D.} 1996 {Data
  assimilation using incremental analysis updates}. {\em Mon. Weather Rev.\/}
  {\bf 124}, 1256--1271.

\bibitem[Bretherton(1970)]{Bretherton70}
{\sc Bretherton, F.~P.} 1970 {A note on Hamilton's principle for perfect
  fluids}. {\em J. Fluid Mech.\/} {\bf 44}, 19--31.

\bibitem[{\c{C}}al{\i}k \& Oliver(2013)]{CalikOliver13}
{\sc {\c{C}}al{\i}k, M. \& Oliver, M.} 2013 Weak solutions for generalized
  large-scale semigeostrophic equations. {\em Commun. Pure Appl. Anal.\/} {\bf
  12}~(2), 939--953.

\bibitem[{\c{C}}al{\i}k {\em et~al.\/}(2013){\c{C}}al{\i}k, Oliver \&
  Vasylkevych]{CalikEtAl13}
{\sc {\c{C}}al{\i}k, M., Oliver, M. \& Vasylkevych, S.} 2013 Global
  well-posedness for the generalized large-scale semigeostrophic equations.
  {\em Arch. Ration. Mech. Anal.\/} {\bf 207}~(3), 969--990.

\bibitem[Cullen(2008)]{cullen2008comparison}
{\sc Cullen, M. J.~P.} 2008 A comparison of numerical solutions to the eady
  frontogenesis problem. {\em Quart. J. R. Meteorol. Soc.\/} {\bf 134}~(637),
  2143--2155.

\bibitem[Cullen \& Purser(1984)]{cullen1984extended}
{\sc Cullen, M. J.~P. \& Purser, R.~J.} 1984 An extended {L}agrangian theory of
  semi-geostrophic frontogenesis. {\em J. Atmos. Sci.\/} {\bf 41}~(9),
  1477--1497.

\bibitem[Eliassen(1948)]{eliassen1948quasi}
{\sc Eliassen, A.} 1948 The quasi-static equations of motion with pressure as
  independent variable. {\em Geofys. Publ.\/} {\bf 17}, 1--44.

\bibitem[Goldstein(1980)]{Goldstein}
{\sc Goldstein, H.} 1980 {\em Classical Mechanics\/}, 2nd edn. Addison-Wesley,
  Reading, MA.

\bibitem[Gottwald(2014)]{Gottwald14}
{\sc Gottwald, G.~A.} 2014 {Controlling balance in an ensemble {K}alman
  filter}. {\em Nonlinear Proc. Geoph.\/} {\bf 21}, 417--426.

\bibitem[Gottwald \& Oliver(2014)]{GottwaldO:2014:SlowDD}
{\sc Gottwald, G.~A. \& Oliver, M.} 2014 Slow dynamics via degenerate
  variational asymptotics. {\em Proc. R. Soc. Lond. Ser. A Math. Phys. Eng.
  Sci.\/} {\bf 470}~(2170), 20140460.

\bibitem[Greybush {\em et~al.\/}(2011)Greybush, Kalnay, Miyoshi, Ide \&
  Hunt]{GreybushEtAl11}
{\sc Greybush, S.~J., Kalnay, E., Miyoshi, T., Ide, K. \& Hunt, B.~R.} 2011
  Balance and ensemble {K}alman filter localization techniques. {\em Mon.
  Weather Rev.\/} {\bf 139}~(2), 511--522.

\bibitem[Hoskins(1975)]{Hoskins75}
{\sc Hoskins, B.~J.} 1975 The geostrophic momentum approximation and the
  semi-geostrophic equations. {\em J. Atmos. Sci.\/} {\bf 32}~(2), 233--242.

\bibitem[Kepert(2009)]{Kepert09}
{\sc Kepert, J.~D.} 2009 {Covariance localisation and balance in an ensemble
  {K}alman filter}. {\em Quart. J. R. Meteorol. Soc.\/} {\bf 135}~(642),
  1157--1176.

\bibitem[Lynch(2006)]{Lynch}
{\sc Lynch, P.} 2006 {\em {The Emergence of Numerical Weather Prediction:
  Richardson's Dream}\/}. Cambridge University Press, Cambridge.

\bibitem[McIntyre \& Roulstone(2002)]{McintyreR2002HigherAA}
{\sc McIntyre, M.~E. \& Roulstone, I.} 2002 Are there higher-accuracy analogues
  of semigeostrophic theory? In {\em Large-scale atmosphere--ocean dynamics\/}
  (ed. J.~Norbury \& I.~Roulstone), vol.~2, pp. 301--364. Cambridge University
  Press.

\bibitem[Mitchell {\em et~al.\/}(2002)Mitchell, Houtekamer \&
  Pellerin]{MitchellEtAl02}
{\sc Mitchell, H.~L., Houtekamer, P.~L. \& Pellerin, G.} 2002 {Ensemble size,
  balance, and model-error representation in an ensemble {K}alman filter}. {\em
  Mon. Weather Rev.\/} {\bf 130}~(11), 2791--2808.

\bibitem[Mohebalhojeh \& Dritschel(2001)]{MohebalhojehDritschel01}
{\sc Mohebalhojeh, A.~R. \& Dritschel, D.~G.} 2001 Hierarchies of balance
  conditions for the {$f$}-plane shallow-water equations. {\em J. Atmos.
  Sci.\/} {\bf 58}~(16), 2411--2426.

\bibitem[Mohebalhojeh \& Dritschel(2004)]{MohebalhojehDritschel04}
{\sc Mohebalhojeh, A.~R. \& Dritschel, D.~G.} 2004 Contour-advective
  semi-{Lagrangian} algorithms for many-layer primitive equation models. {\em
  Quart. J. R. Meteorol. Soc.\/} {\bf 130}, 347--364.

\bibitem[Oliver(2006)]{Oliver06}
{\sc Oliver, M.} 2006 Variational asymptotics for rotating shallow water near
  geostrophy: a transformational approach. {\em J. Fluid Mech.\/} {\bf 551},
  197--234.

\bibitem[Oliver \& Vasylkevych(2011)]{OliverVasylkevych11}
{\sc Oliver, M. \& Vasylkevych, S.} 2011 Hamiltonian formalism for models of
  rotating shallow water in semigeostrophic scaling. {\em Discrete Contin. Dyn.
  Syst.\/} {\bf 31}~(3), 827--846.

\bibitem[Oliver \& Vasylkevych(2013)]{Oliver:2013}
{\sc Oliver, M. \& Vasylkevych, S.} 2013 Generalized {LSG} models with
  spatially varying {C}oriolis parameter. {\em Geophys. Astrophys. Fluid
  Dyn.\/} {\bf 107}, 259--276.

\bibitem[Ourmi\'eres {\em et~al.\/}(2006)Ourmi\'eres, Brankart, Berline,
  Brasseur \& Verron]{OurmieresEtAl06}
{\sc Ourmi\'eres, Y., Brankart, J.~M., Berline, L., Brasseur, P. \& Verron, J.}
  2006 {Incremental analysis update implementation into a sequential ocean data
  assimilation system}. {\em J. Atmos. Ocean. Tech.\/} {\bf 23}, 1729--1744.

\bibitem[Richardson(1922)]{Richardson22}
{\sc Richardson, L.~F.} 1922 {\em Weather Prediction by Numerical Process\/}.
  Cambridge University Press, Cambridge.

\bibitem[Salmon(1983)]{Salmon83}
{\sc Salmon, R.} 1983 Practical use of {H}amilton's principle. {\em J. Fluid
  Mech.\/} {\bf 132}, 431--444.

\bibitem[Salmon(1985)]{Salmon1985}
{\sc Salmon, R.} 1985 New equations for nearly geostrophic flow. {\em J. Fluid
  Mech.\/} {\bf 153}, 461--477.

\bibitem[Salmon(1998)]{Salmon}
{\sc Salmon, R.} 1998 {\em {Lectures on Geophysical Fluid Dynamics}\/}. Oxford
  University Press, New York.

\bibitem[Smith \& Dritschel(2006)]{SmithDritschel:2006}
{\sc Smith, R.~K. \& Dritschel, D.~G.} 2006 {Revisiting the
  {R}ossby--{H}aurwitz wave test case with contour advection}. {\em J. Comput.
  Phys.\/} {\bf 217}~(2), 473--484.

\bibitem[Vi{\'u}dez \& Dritschel(2004)]{ViudezDritschel04}
{\sc Vi{\'u}dez, {\'A}. \& Dritschel, D.~G.} 2004 Optimal potential vorticity
  balance of geophysical flows. {\em J. Fluid Mech.\/} {\bf 521}, 343--352.

\end{thebibliography}


\end{document}